\documentclass[conference]{IEEEtran}

\usepackage{enumitem}
\usepackage{array}
\usepackage{amsmath}
\usepackage{graphicx}
\usepackage{mathtools}
\usepackage{cuted}
\usepackage{lipsum, color}
\usepackage{subfigure}

\usepackage{tikzpagenodes} 

\usepackage{caption} 
\graphicspath{ {Figs/} }

\newcolumntype{+}{>{\global\let\currentrowstyle\relax}}
\newcolumntype{^}{>{\currentrowstyle}}
\newcommand{\rowstyle}[1]{\gdef\currentrowstyle{#1}%
	#1\ignorespaces
}

\begin{document}

\title{An Empirical Evaluation Of Social Influence Metrics}

\author{\IEEEauthorblockN{ Nikhil Kumar, Ruocheng Guo, Ashkan Aleali, Paulo Shakarian }
	\IEEEauthorblockA{Arizona State University,\\
		Tempe, AZ\\
		Email: \{nikhilkumar, rguosni, aleali, shak\}@asu.edu}}

\maketitle  

\begin{abstract}
Predicting when an individual will adopt a new behavior is an important problem in application domains such as marketing and public health.  This paper examines the performance of a wide variety of social network based measurements proposed in the literature - which have not  been previously compared directly.  We  study the probability of an individual becoming influenced based on measurements derived from neighborhood (i.e. number of influencers, personal network exposure), structural diversity, locality, temporal measures, cascade measures, and metadata.  We also examine the ability to predict influence based on choice of classifier and how the ratio of positive to negative samples in both training and testing affect prediction results - further enabling practical use of these concepts for social influence applications.
\end{abstract}

\section{Introduction}  

Predicting when an individual will adopt a new behavior is an important problem in application domains such as marketing~\cite{watts07}, the spread of innovation~\cite{val95}, countering extremism~\cite{ag15}, and public health~\cite{centola10}.  As a result, a variety of social network based measurements have been proposed in the literature and shown to predict how likely an individual will adopt a new behavior given information about his immediate social ties.  However, when such measures are proposed, they are often evaluated under different conditions - making it difficult to understand which of these measurements should be used in a real-world application.  Further complicating the issue is that the choice of classification algorithm and the effect of class imbalance in both training and testing are often not explored in most research. In our lab, we have the goal of creating and deploying a system for counter-extremism messaging.  Hence, understanding how influence measurements work in experimental settings that closely resemble real-world scenarios is an important first step.

In this paper, we study measurements based on neighborhood (i.e. number of influencers~\cite{centola10}, personal network exposure~\cite{val95}), structural diversity~\cite{ugander12}, locality~\cite{zhang2013social}, temporal measures~\cite{goyal2010learning}, cascade measures~\cite{guo2015toward}, and metadata~\cite{jenders2013analyzing}.  We examine the probability of an individual becoming influenced based on these measurements (probability of adoption).  We also examine the ability to predict influence based on choice of classifier and how the ratio of positive to negative samples in both training and testing affect prediction results.  Specifically, we make the following contributions.
\begin{enumerate}
	\item\label{it1}{We review a variety of measurements used to predict social influence and we group them in six categories (Section~\ref{features}).}
	\item\label{it2}{We evaluate how  these measurements  relate to the probability of a user being influenced using real-world microblog data (Section~\ref{measStudy}).}
	\item\label{it3}{We evaluate how these measurements perform when used as features in a machine learning approach and compare performance across a variety of supervised machine learning approaches (Section~\ref{infpred}).}
	\item\label{it4}{We evaluate how the ratio of positive to negative samples in both training  and testing affect predictive results (Section~\ref{imbalSec}).}
\end{enumerate}
We note that contribution~\ref{it4} is of particular importance, as (particularly with microblog data) users are exposed to large number of messages that they do not retweet (negative samples).  Hence, in both training and testing, researchers can increase the negative samples utilized by large amounts - hence arbitrarily determining the level of class imbalance. As with this study as a whole, the experiments on data imbalance were to better understand these previous research results in tests that better mimicked real-world scenarios.

\vspace{8pt}
\noindent\textbf{Related work.}  Beyond the work that we shall describe concerning the various measures for social influence we investigate in Section~\ref{measStudy}, there has been some general work in the area of social influence that have taken approaches not necessarily amenable to comparison.  For instance, the seminal work of Kempe et al. \cite{kempe2003maximizing} describes two popular models for information cascades which spawned several techniques to learn the parameters (which also correspond to edge weights in the graph).  For example, Saito et al. \cite{saito2008prediction} assigned such probabilities based on an expectation-maximization appproach while Goyal et al. ~\cite{goyal2010learning} leveraged a variety of simple models based on ideas such as a empirically-learned probabilities and similarity measurements.  See ~\cite{shakarian2015diffusion} for a review of some of this work.  There has also been related work on predicting cascades~\cite{cheng2014can,weng2013virality,guo2015toward} which are more focused on determining if a trend in social media exceeds a certain size.  That said, some of the ideas from these approaches, such as structural diversity~ \cite{ugander12} are examined here (though this paper is focused on a different problem).  Other work such as Myers et al. \cite{myers2012information} studied the external factors influencing information diffusion, Liu et al. \cite{liu2010mining} and Tang et al. \cite{tang2009social} focused their studies on topic influence. Jenders et al.~\cite{jenders2013analyzing} studied a combination of different features including some of the metadata features like mentions and hashtags, along with latent features like sentiments and emotional divergence for predicting the virality of a tweet - many of which we examine in this study as well. Hong et al.~\cite{hong2011predicting} have also considered a wide spectrum of features including structural, content and temporal information. However, their study focused more on content-based features and not the structural features considered here - many of which were introduced after that work.

\section{Technical Preliminaries} \label{sec:tech_pre}

Here we introduce the necessary notation and describe our social network data.  We represent a social network as a graph $G=(V,E)$ where $V$ is the set of vertices and $E$ is the set of directed edges that have sizes $|V|,|E|$ respectively. The intuition behind edge $(v,v')$ is that node $v$ can influence $v'$. This intuition stems from how we create the edges in our network: $(v,v')$ is an edge if during a specified time period there is at least one microblog posted by $v$ that is reposted by $v'$ . For node $v \in V$ , the set of in-neighbors is denoted as $\eta_{v}^{in}$ , and the set of out-neighbors as  $\eta_{v}^{out}$ . We use   $d_{v}^{in}$ and $d_{v}^{out}$ to denote the in-degree and out-degree respectively.  We also assume a partition over nodes that specifies a community structure.  We assume that such a partition is static (based on the same time period from which the edges were derived) and the function $P(V):V \rightarrow \mathcal{C}$ maps the set of nodes ($V$) to the set of communities ($\mathcal{C}$), where $\mathcal{C}$ consists of $k$ communities: $\{C_1, C_2, ...,C_k \}$. We utilize the Louvain algorithm \cite{blondel2008fast} to identify our communities in this paper due to its ability to scale.\\

\noindent\textbf{Cascades.} For a given microblog $\theta$, we define $t$ as the number of time units from the initial post of $\theta$ before the microblog was reposted by one of $v$'s incoming neighbors - intuitively the time at which $v$ was exposed to $\theta$. We denote the subset of nodes who originally posted or reposted $\theta$ for time period $t$ as $V_{\theta}^t$.  Likewise, the set of reposting relationships within the same time period will be denoted by $R_{\theta}^t$.  Taken together, we have a \textit{cascade}: $D_{\theta}^t = (V_{\theta}^t,R_{\theta}^t)$.  Any valid original microblog $\theta$ could be treated as a unique identifier for a cascade. Given a microblog $ \theta $, $v_{\theta}$ is the originator at instance $t_{\theta}^0$, which is defined as the origin time when the originator posted the microblog $\theta$.  We denote the size of a cascade at any particular time $t$ as $|V_{\theta}^t|$.  For $v \in V_{\theta}^{t} $, the set of all \textit{active} neighbors with respect to $\theta$ is defined as $S_{\theta}^{v} = V_{\theta}^{t} \cap \eta_{v}^{in}$ . We also define the distance $d_{\theta}^{t}(v,u)$ as the shortest path length between $v$ and $u$ in $D_{\theta}^t$.  \\

\noindent \textbf{Sina Weibo Dataset.} The dataset we used was provided by the WISE 2012 Challenge\footnote{http://www.wise2012.cs.ucy.ac.cy/challenge.html}. 
It included a sample of microblogs posted on Sina Weibo from 2009 to 2012. In this dataset, we are provided with time and user information for each post and the last repost in a chain which enabled us to derive a corpus of cascades. We create the social network $G$ from the retweeting relationships of microblogs published between May 1, 2011 and July 31, 2011. We use the microblogs published in August 2011 to train and test our approach.   Table~\ref{tab:ds} lists the statistics of the dataset we used.

\begin{table}[!h]
	\centering
	\small
	\begin{tabular}{|c|c|c|c|p{1.4cm}} \hline
		\#Users & \#Edges & \#Reposted tweets &  \#Reposted Users\\ \hline
		5,910,608 & 52,472,547 & 2,238,659 & 394,441\\ \hline
	\end{tabular}
	\caption{Graph statistics}
	\label{tab:ds}
\end{table}

We found that the network derived from the dataset had 7,668,693 users with 55,381,104 edges between them. For this network, the number of active users in August (the time period used to study social influence) is 5,910,608 while 5,664,625 of them have at least have one out-neighbor. During the month of August, there were 22,182,703 retweet chains.  From this data, we removed the users who are not present in $V$; we also removed 2,660,421 empty repost chains caused by this elimination.  The dataset does not contain the repost time for the nodes in the middle of chains. We estimated this time for each node in the chain based on the original post time and the final repost time.  Table~\ref{tab:ds} lists the statistics of this dataset during the period of study.

Among all the retweeted users we further extract the top retweeters defined as those who had at least 100 retweets during the period. This set of high frequency tweeters will be used as a base for deriving the sample set for our experiments. For each user in the above mentioned group, an occurrence of them retweeting a post when they have an active in-neighbor is considered as a positive instance. If any of their followees have tweeted and they haven't retweeted, it is considered as a negative instance. 

\section{Measurements to Predict Social Influence} \label{features}
In this section, we categorize several approaches for predicting social influence.  
\begin{enumerate}
	\item Neighborhood-based measures
	\item Structural diversity measures
	\item Influence locality
	\item Cascade-based measures
	\item Temporal measures
	\item Metadata
\end{enumerate}
We examine each of these categories in turn.

\vspace{5pt}

\noindent \textbf{Neighborhood-based measures.} These are the measures computed using each node and its immediate neighbors. These measures represents the pair wise influence that the neighboring nodes exert on a given node. Retweeting from followees is the primary mode of tweet visibility in a microblogging site, as usually a tweet is visible to a user from its followee subgraph.  Specifically, we study the following

\begin{itemize}
	\item \textbf {Number of active neighbors.}  ($|{S_{v}^{\theta}}|$)  This represents the count of active neighbors for a node $v$.  In Damon Centola's notable empirical study~\cite{centola10}, he noted that additional ``social signals'' -- or active neighbors -- significantly increased the likelihood of an individual adopting a new behavior.
	
	\item \textbf{Personal Network Exposure (PNE).} ($|{S_{v}^{\theta}}| / d_{v}^{in}$) Is a measure adopted from the social science community (i.e. see \cite{val95} ) and has obtained recent interest (i.e. \cite{hal14}). As per \cite{val95}, PNE quantifies the extent to which a person is exposed to direct and indirect influence. This value is defined as the ratio of number of active neighbors to total number of neighbors. It is a measure of the fraction of influence an active neighbor $u$ has on $v$. If $v$ has many in-neighbors aka followees, then $u$'s influence is diluted and PNE represents that dilution.
	
	\item \textbf{Average in-neighbor count of active neighbors.} ($ |{ \Sigma_{u \in  S_{v}^{\theta} } d_{u}^{in} }| / |{S_{v}^{\theta}}|  $)
	This is calculated by averaging the number of in-neighbors of each active neighbor of a node. This defines the dilution of the influence path and is similar to the measure, \textit{number of uninfected neighbors} as described in ~\cite{weng2013virality}. Other releated studies include Cha et al. ~\cite{cha2010measuring}, where they studied the effect of a social network user's indegree in depth, and observed that high indegree is not necessarily correlated to influence in terms of spawning retweets.   
	
\end{itemize}

\noindent  \textbf{Structural diversity measures.} This group of measurements taken into account the structural diversity in the local neighborhood of the node - which refers to the communities present in the neighborhood.

Ugander et al. ~\cite{ugander12} introduced structural diversity where they studied the effect of number  of connected components of a friendship network. Fortunato et al \cite{fortunato2010community} defined communities as the set of graph vertices which are organized into groups that seem to live fairly independently of the rest of the graph. Weng et al. ~\cite{weng2014predicting} used the community structure to predict the increase in cascade size. We use the modularity maximization method ~\cite{chen2014community} for detecting communities in our dataset. The Louvain Algorithm ~\cite{blondel2008fast} which comes under this method is used to derive the communities in this study due to its ability to scale. We use two community based measures. 
\begin{itemize}	
	\item \textbf{Active community count.}  ($|P(S_{v}^{\theta})|$)
	This is defined as the number of adjacent communities of a given user $v$ with at least one active neighbor of $v$. The communities that include active neighbors are more significant in this context than rest of the adjacent communities. Shakarian et al. have studied this measure in their book ~\cite{shakarian2015diffusion} highlighting the importance of structural diversity.
	
	\item \textbf{Active community ratio} ($|P(S_{v}^{\theta})|/|P(\eta_{v}^{in})|$) 
	It is calculated as the ratio of the active community count to the total number of adjacent communities. This is similar to the personal network exposure \cite{val95} and represents the dilution of the effect of active community count with respect to other neighboring communities. 
\end{itemize}

\noindent\textbf{Influence locality.}  We examine the Influence Locality model known as LRC-Q, introduced by Zhang et al. \cite{zhang2013social}. LRC-Q is defined by the influence locality function $Q$ which is a combination of peer influence factor ($g$) and structural factor ($f$). Peer influence factor is obtained as a linear combination of the geometric mean of random walk probabilities of active neighbors and structural factor as a linear combination of the number of circles formed by the active neighbors in the ego network of the user $v$. These are defined in their paper by the following equations.
\begin{equation}
	Q = w \times g + (1-w) \times f
\end{equation}
\begin{equation}
	g = \sqrt[\leftroot{-3}\uproot{6}|{S_{v}^{\theta}}|]{\prod_{v_{i} \in S_{v}^{\theta}}  (t_{\theta}^{v} - t_{\theta}^{v_{i}}) \times p_{v_{i}}}
\end{equation}
\begin{equation}
	f=a\log(|{S_{v}^{\theta}}|+1) + be^{-\mu|C(S_{v}^{\theta})|}
\end{equation}
In the above equations, $p_{v_{i}}$ is the random walk probability from the active user $v_{i}$ to the given user $v$, $C(S_{v}) $ is the collection of circles formed by the active neighbors, $t_{\theta}^v$ is the time at which $v$ posted or reposted the microblog $\theta$, $\mu$ is the decay factor and, $a$, $b$ and $w$ are balance parameters. For our experiments we set the value of $\mu$ as 1 and, $a$, $b$ and $w$ to be 0.5, as per the parameter settings of \cite{zhang2013social}. 

\noindent \textbf{Cascade-based measures.} 

This group of measurements take into account the various parameters that are part of a microblog cascade. There has been many studies in the area of predicting the cascades including Bakshy et al. ~\cite{bakshy2011everyone} , Cheng et al. ~\cite{cheng2014can} and more recently Guo et al. ~\cite{guo2015toward}. Unlike our study, there hasn't been many attempts to utilize the cascade parameters in predicting retweet behavior. We study the following measures.

\begin{itemize}
	
	\item \textbf{Cascade size.} ($|V_\theta^t|$)
	Cascade size is computed as the count of people who have retweeted a particular microblog $\theta$ at time $t$. This number is usually visible to the microblog user and can have an impact on their retweet behavior.

	\item \textbf{Path length.} ($d_{\theta}^{t}(v, v_{\theta}) $)
	Path length is the length of a tweet trace path from the original tweeter to a given user in the cascade. Watts et al. ~\cite{watts1998collective} were the first to study the path length where they found that many social and technological networks have small path lengths. Kwak et al. ~\cite{kwak2010twitter} studied the path length in twitter, and Weng et al. ~\cite{weng2014predicting} studied a distance measure called Average step distance which was based on the path length. Our study focuses on the path length with respect to a particular cascade $D_{\theta}^t$.\\

\end{itemize}

\noindent \textbf{Temporal Measure}
Temporal measures were given prominence in many of the prior studies either by itself, or as a factor in combination with other measures. Goyal et al. ~\cite{goyal2010learning} utilized the temporal factor and attempted to predict the time by which an influenced user will perform an action. Hong et al. ~\cite{hong2011predicting} studied a variety of temporal measures and observed that they have a stronger effect on messages with low and medium volume of retweets, compared to highly popular messages. We study the following temporal measure.

\begin{itemize}
	\item \textbf{Retweet Time delay.} ($t$)
	This is defined as the time delay between the original tweet and the time when  $v$ is exposed to microblog $\theta$. The time at which a tweet was made  is another piece of information which people are exposed to while viewing a tweet.  This can affect their decision to retweet it or not. This is one of the temporal measures studied by Hong et al. ~\cite{hong2011predicting}.\\
	
\end{itemize}

\noindent \textbf{Metadata.} These are simple measures derived from the metadata associated with the tweets. We consider the presence or absence of links, mentions and hashtags as measures for our study.  Jenders et al.~\cite{jenders2013analyzing} did an extensive analysis of a wide range of tweet and user features regarding their influence on the spread of tweets. They considered the number of mentions and number of hashtags among the obvious tweet features. They observed that tweets containing  both hashtags and mentions are more likely to be retweeted than those with out, however as the number of hashtags/mentions in a tweet grows, the expected number of retweets decreases. In this study we only consider their presence or absence as a measure and do not go into any deeper analysis. 

\begin{itemize}
	\item \textbf{Presence of a link (hasLink).}  This is a binary value which represents whether the original tweet had a link. Links are usually shown as part of the tweet content. The measure of Links in tweets is similar to that of mentions and hashtags, but has not been studied as extensively as either in the context of social influence.
	
	\item \textbf{Presence of a mention (hasMention).}
	A binary value which represents whether the original tweet had a mention. Intuitively, a user might be more willing to retweet if there is a mention of him/her or someone he/she knows. Similar to ~\cite{jenders2013analyzing},  Cha et al. ~\cite{cha2010measuring} analyzed the effect of the number of mentions and found that mentions can be an important measure of an individual influence in the social network. 
	
	\item \textbf{Presence of a hashtag (hasHashtag).}
	A binary value which represents whether the original tweet had hashtags. Hashtags are also means by which tweets become visible to users and thus are of significance in this regard. A deeper analysis such as ~\cite{jenders2013analyzing}, is beyond the scope of this work and we only focus on how the presence or absence of a hashtag affects the retweeting behavior. \\
\end{itemize}

\section{Social Influence Measurement Study}
\label{measStudy}

Here, we examine the distribution of various measurements which were defined in Section~\ref{features}. For each of those measures, the values are put into intervals of equal sizes and the fraction of positive samples in the interval is plotted as the probability. The horizontal axis shows the value intervals of the measure, while the vertical one shows the number of occurrences for positive instances with respect to the total amount in that particular interval. The error bar shows twice the standard deviation of the sample. These are shown in Fig.~\ref{fig:neighb_temp} and Fig.~\ref{fig:struct_nw}. A detailed analysis of their distribution is given below. \\

\noindent \textbf{Neighborhood-based measures.}
Active neighbor count intuitively has a positive correlation with the influence as shown in Fig.~\ref{fig:active_neighb}. Fig.~\ref{fig:active_neighb9} shows the active neighbor count for the lower values which also shows similar correlation. This is consistent with the empirical study of ~\cite{centola10}. As the number of retweeters among in-neighbors increases, the probability of a person retweeting the particular tweet increases. Fig.~\ref{fig:pne} shows that PNE also exhibits positive correlation like active neighbor count. This shows the significance of PNE measure as demonstrated by other studies such as \cite{val95} and  \cite{hal14}. Average in-neighbor count of Active Neighbors does not show a clear correlation in its distribution as seen in Fig. ~\ref{fig:AvgOut}.  \\

\noindent \textbf{Structural diversity measures.}
Number of active communities shows a good positive correlation with the retweet behavior. This result is consistent with the related studies such as ~\cite{weng2014predicting} and ~\cite{cheng2014can}. Active community ratio also demonstrates a reasonable correlation with the positive instances as this measure represents the dilution of community influence based on the total number of adjacent communities. \\

\noindent \textbf{Cascade-based measures.}
Intuitively, cascade size is an important influencer in retweet behavior. If a tweet is reasonably popular it tends to attract further retweets. The same is revealed from the distribution in Fig.~\ref{fig:Cascadesize}. This is consistent with the research of ~\cite{bakshy2011everyone} and ~\cite{cheng2014can}  although they studied a different problem.  The intuition for path length is that, as the distance from the original tweeter increases a user is less interested in retweeting the tweet. Our results show (Fig.~\ref{fig:Pathlength}) that this intuition holds between path length 1 and 2. But, for the remaining intervals, results doesn't correlate well. This can be explained by comparing to the results of ~\cite{jenders2013analyzing} where they found similiar pattern while analyzing mentions and hashtags.  Further, the results of ~\cite{cheng2014can} indicate that information cascade depth is related to popularity.  Hence, the microblogs that are far from the original poster may be inherently popular as the information cascade has proceeded to a larger depth. \\

\noindent \textbf{Temporal.}
Fig.~\ref{fig:rtd} shows that retweet time delay has slight inverse correlation with the influence. Intuitively, the influence of a tweet decays with time, and as people are exposed to date/time information in the social network they are less likely to retweet old tweets. This decay factor has been used in works like ~\cite{goyal2010learning}, ~\cite{zhang2013social} etc. and above result shows the same.  \\

\noindent \textbf{Metadata.}
Table~\ref{tab:meta_feature} shows the conditional probability of positive instances given the meta measure value of 0 and 1, respectively. The values from the table shows that presence or absence of a link doesn't seem to have much correlation with the influence. It also shows that, the presence of mentions seem  have slight negative correlation to influence though there is no actual intuition to base this on. But, this can be explained by the observation in the paper ~\cite{jenders2013analyzing} that as the number of mentions in a tweet grows, the expected number of retweets decreases. The presence of hashtag shows an interesting correlation in Table~\ref{tab:meta_feature}.  This is consistent with the study of ~\cite{jenders2013analyzing} and  illustrates the significance of hashtags in enhancing the visibility of the tweet and motivating a user to retweet them. \\

\begin{table}[!h]
	\centering
	\begin{tabular}{|+p{0.18\columnwidth}|^p{0.33\columnwidth}|+p{0.33\columnwidth}|} \hline 
		\rowstyle{\bfseries}%
		
		$\vec{V}$ & $P(y_{i}= $ pos $| V_{i}=0)$ & $P(y_{i}=$ \textbf{pos} $| V_{i}=1)$   \\  \hline \hline
		hasLink & 0.51 & 0.48   \\ \hline
		hasMention & 0.51 & 0.45   \\ \hline
		hasHashtag & 0.50 & 0.66  \\ \hline
		
	\end{tabular}
	\caption{$\vec{V}$ is a column of the design matrix corresponding to a certain binary feature, \textbf{pos} represents positive label and $i$ is the index of the sample.}
	\label{tab:meta_feature}
\end{table}

\begin{figure}[!h]  
	\centering
	\subfigure[]{
		\includegraphics[width=.46\columnwidth]{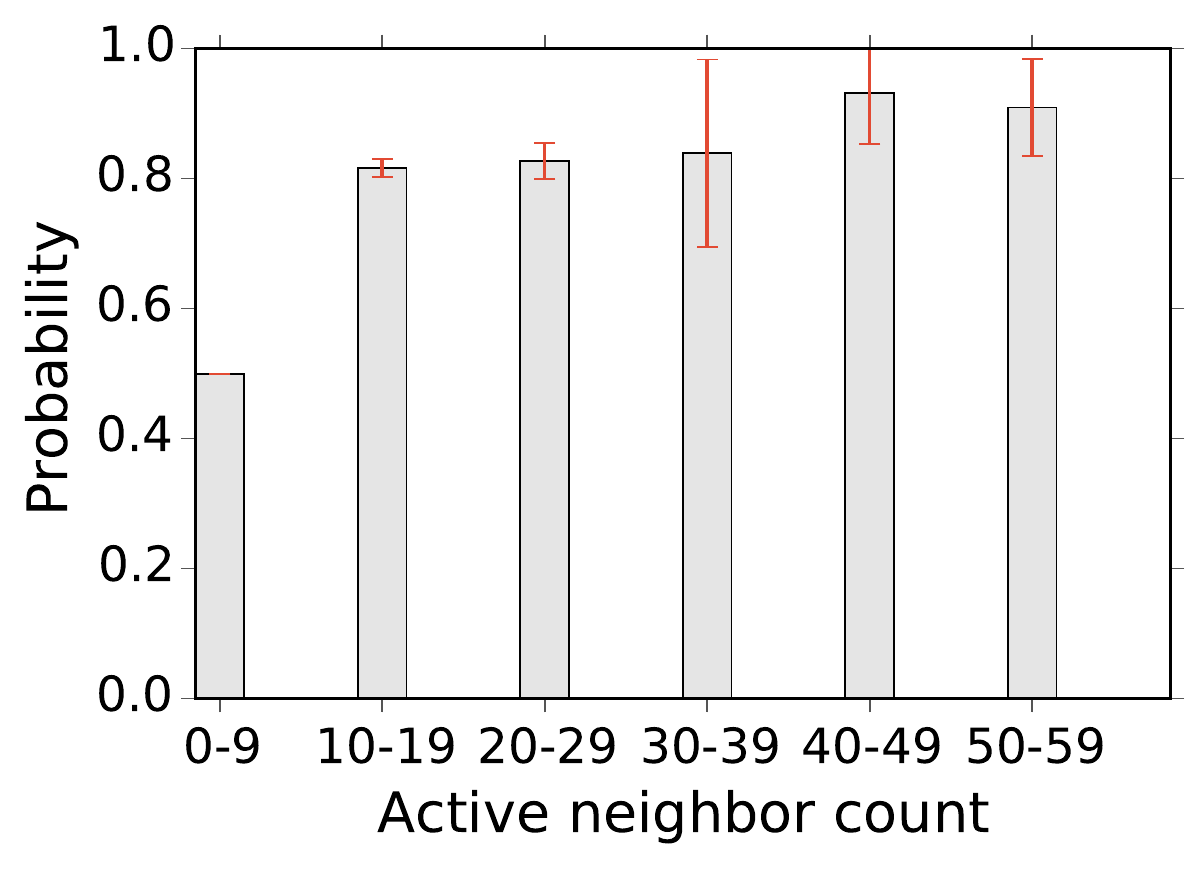}
		\label{fig:active_neighb}
	} 
	\subfigure[]{
		\includegraphics[width=.46\columnwidth]{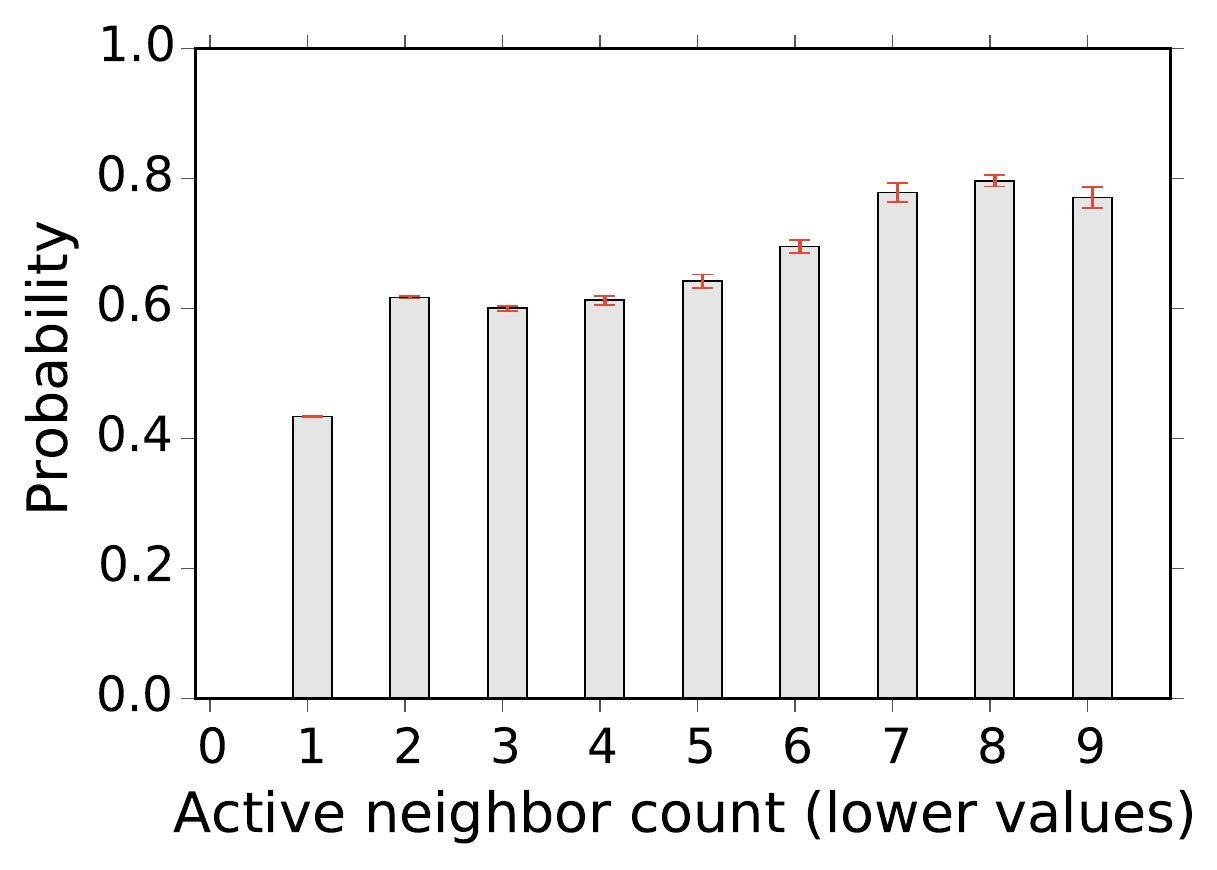}
		\label{fig:active_neighb9}
	} 
	\subfigure[]{%
		\includegraphics[width=.47\columnwidth]{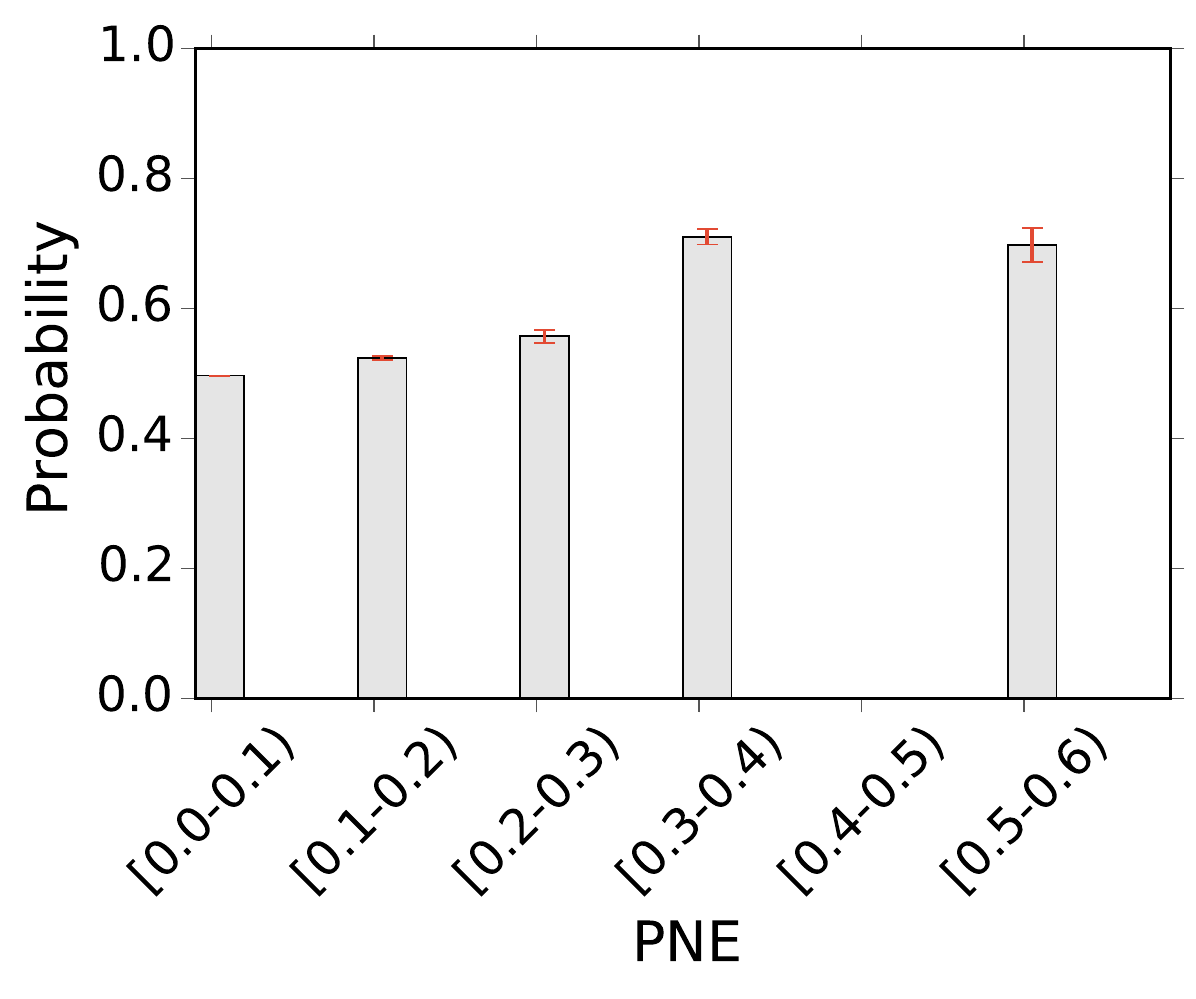}
		\label{fig:pne}
	} 
	\subfigure[]{%
		\includegraphics[width=.47\columnwidth]{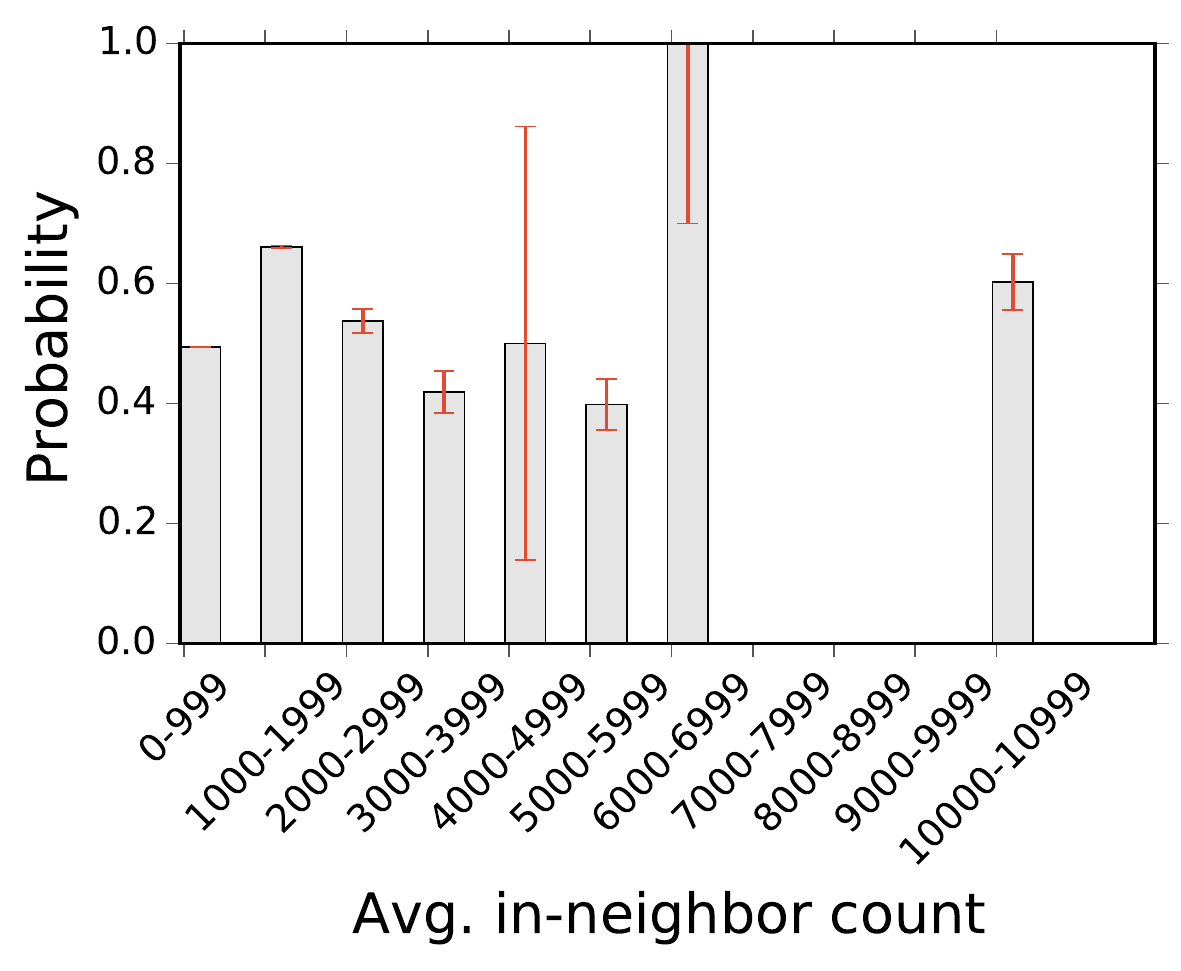}
		\label{fig:AvgOut}
	}    
	\caption{Plots of Neighborhood and temporal measures. Error bars represent two standard deviations.}
	\label{fig:neighb_temp}
\end{figure}

\begin{figure}[!h]
	\centering
	\subfigure[]{%
		\raisebox{.14\height}{ 
			\includegraphics[width=.47\columnwidth]{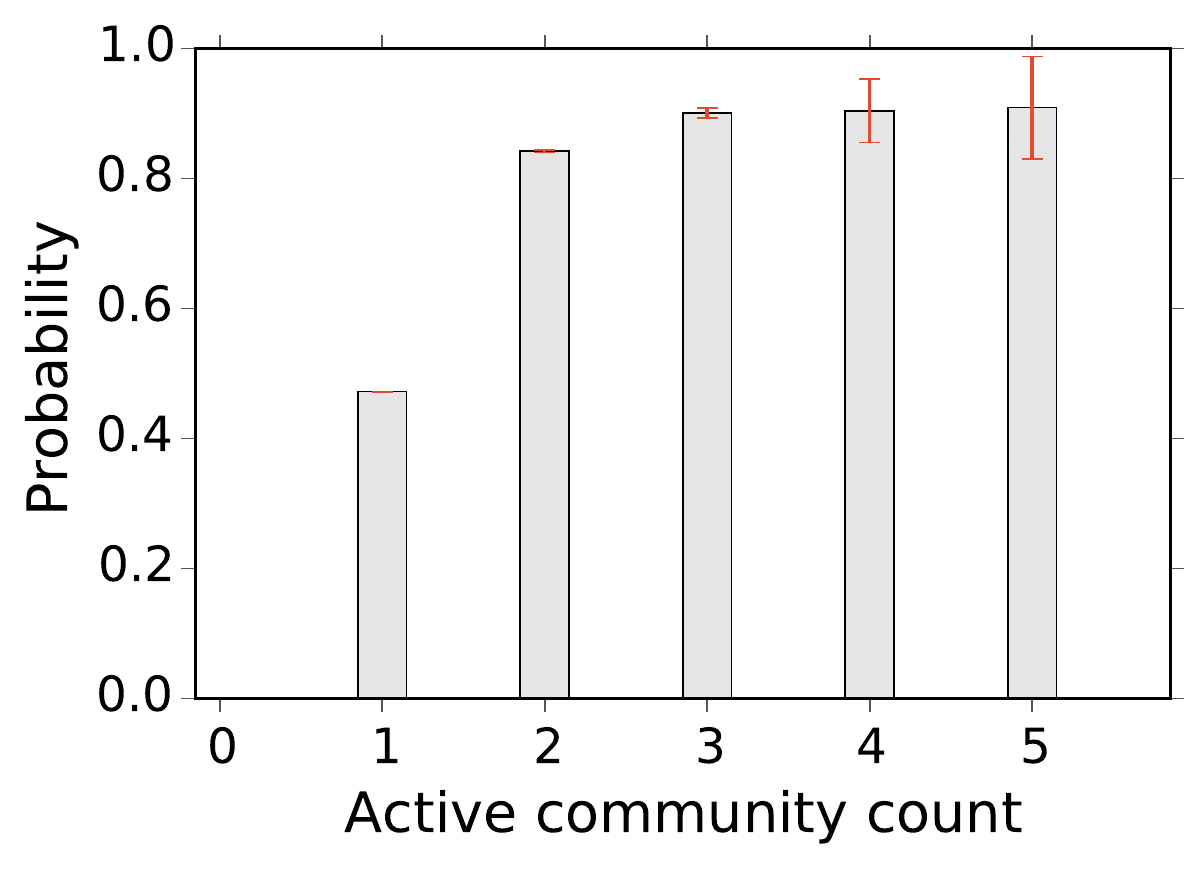}}
		\label{fig:comcnt}
	}
	\subfigure[]{%
		\includegraphics[width=.47\columnwidth]{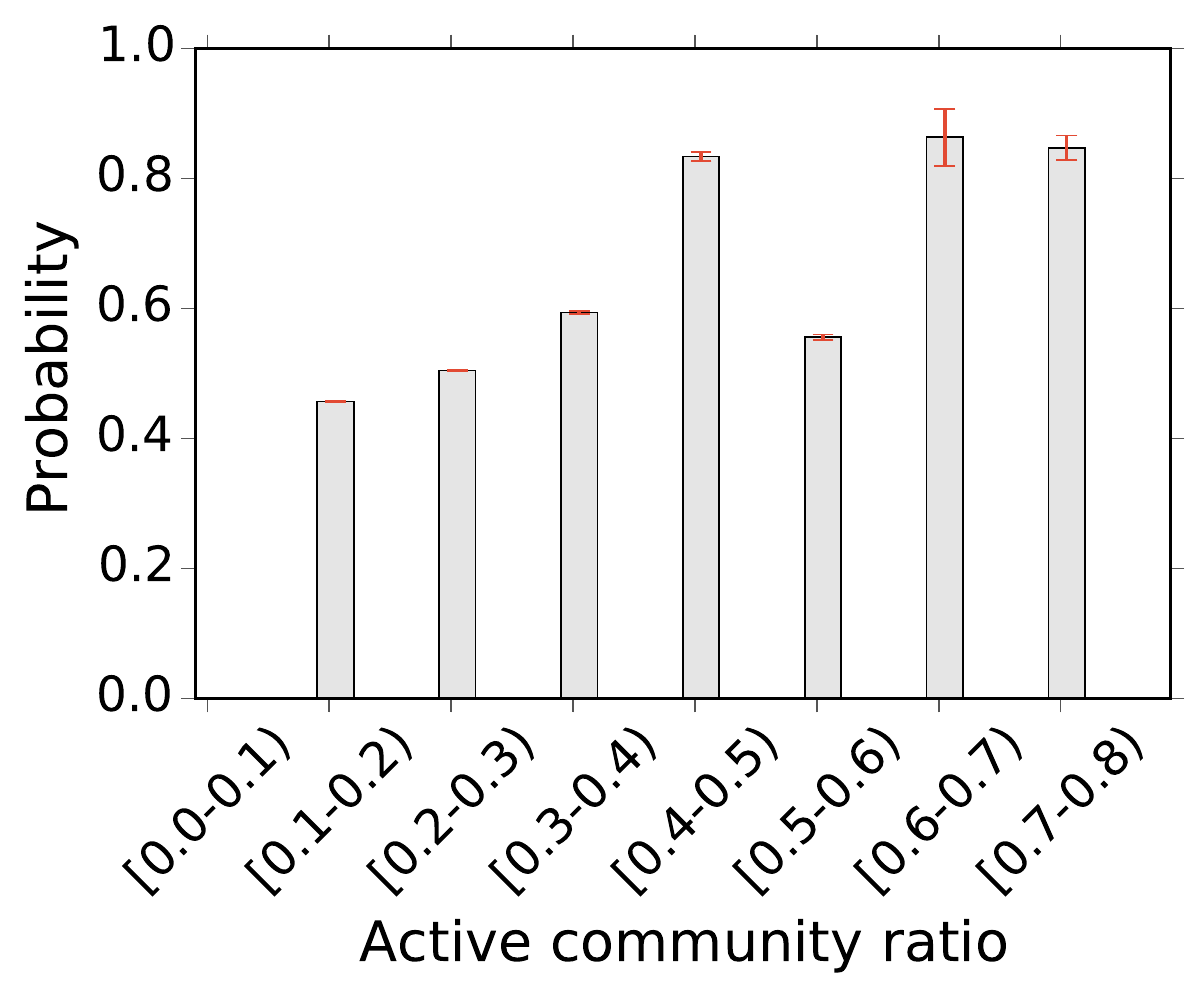}
		\label{fig:CommunityRatio}
	} 
	\subfigure[]{%
		\includegraphics[width=.47\columnwidth]{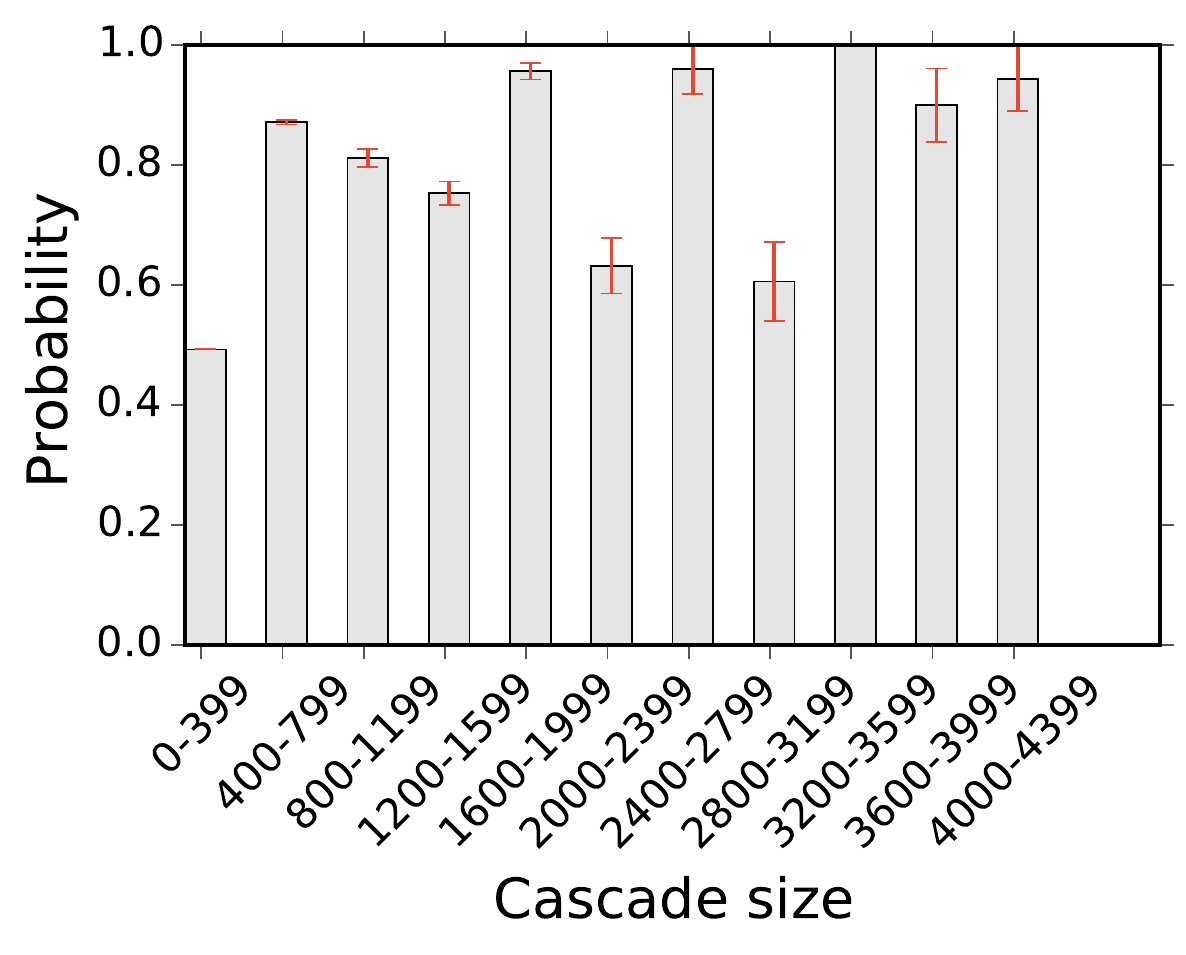} 
		\label{fig:Cascadesize}
	} 
	\subfigure[]{%
		\raisebox{.14\height}{ 
			\includegraphics[width=.47\columnwidth]{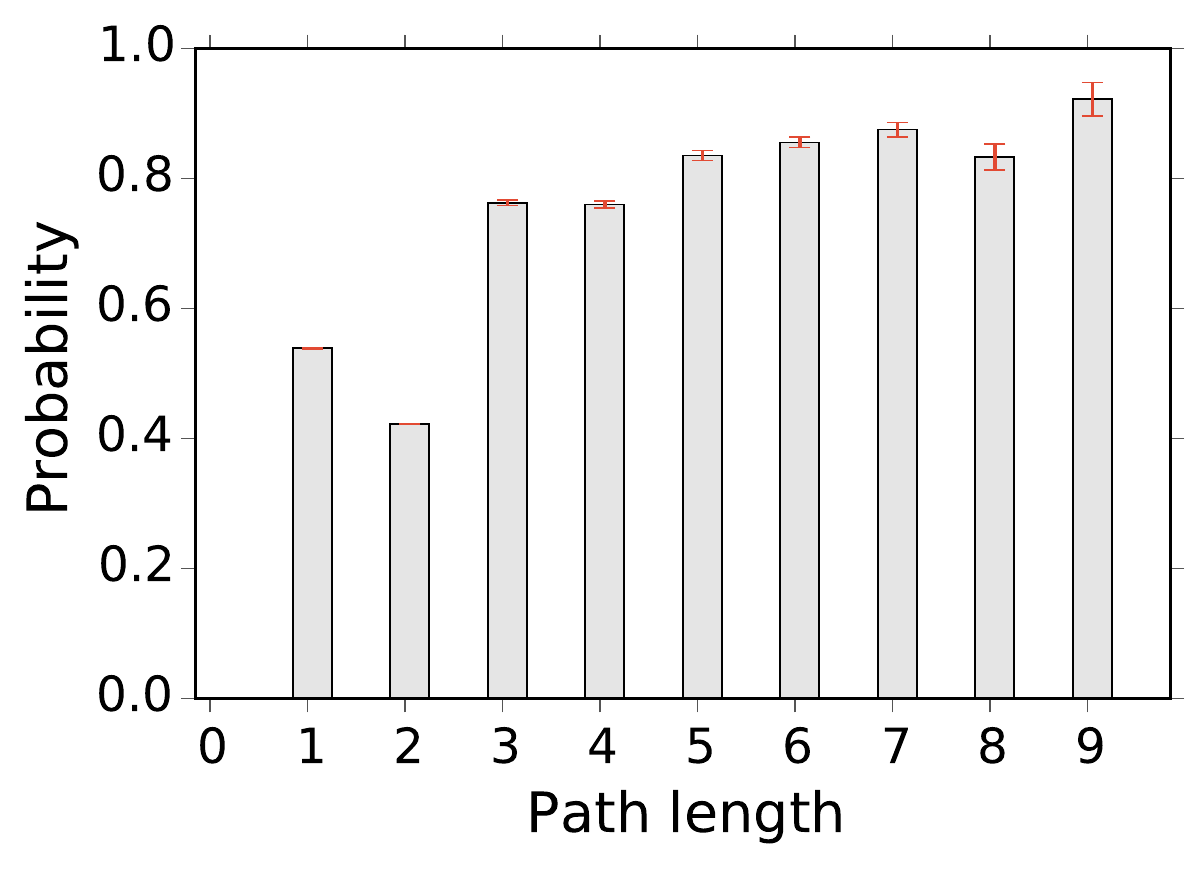}} 
		\label{fig:Pathlength}
	}  
	\subfigure[]{%
		\includegraphics[width=.47\columnwidth]{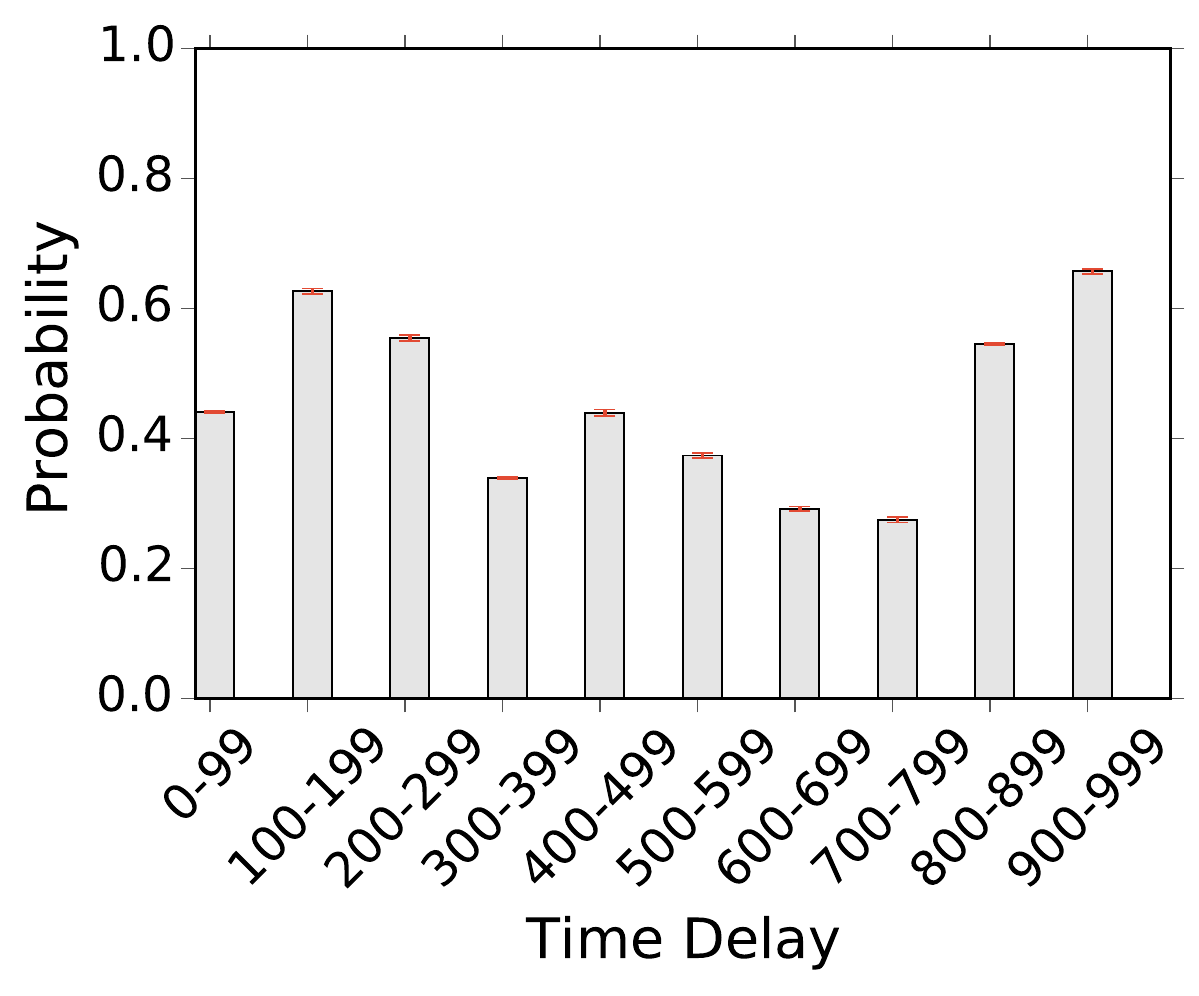}
		\label{fig:rtd}
	}    
	\caption{Plots of Structural Diversity and Cascade-based measures. Error bars represent two standard deviations.}
	\label{fig:struct_nw}
\end{figure}

\section{Influence Prediction}
\label{infpred}

\subsection{Methods}

We derive our graph $G$ from the dataset as described under Section~\ref{sec:tech_pre}. We use the microblogs published in August 2011 to extract the instances to train and test our approach. Positive and negative instances are extracted as described in Section~\ref{sec:tech_pre}, and the measures described in Section  ~\ref{features} were extracted as features for each of them. This set is used to obtain a random sample with 1:1 negative to positive ratio, which we will use for the classification experiments. \\

\noindent  \textbf{Classification experiments}
Here we examine our experiments for predicting whether a user under given conditions will retweet or not. As this is a binary classification task we report the performance measurements (precision, recall and unbiased F1) for only the positive (retweeting) class. We also examine the classification performances of various learning algorithms. For each of the experiments we use a training to test set ratio of 70:30 and used a 10 fold cross validation. We use the following classification algorithms for our experiment.

\vspace{5pt}

\noindent  \textbf{Random Forest (RF)}.
Random Forest \cite{breiman2001random} is a popular ensemble method used for classification and regression. Ensemble methods use multiple classifier algorithms to obtain better accuracy than that could be obtained using any of the individual classifiers. We use random forest algorithm with bootstrap aggregating, that fits a number of decision trees on different sub-samples of the dataset. Each decision tree provides its own predictions which are then merged obtain a better accuracy. \\

\noindent  \textbf{AdaBoost Classifier (AB)}. The AdaBoost algorithm ~\cite{freund1999short} proposed by Yoav Freund and Robert Schapire is one of the most important ensemble methods. It is prominent among the boosting techniques ~\cite{freund1999short} which are used in conjuction with other learning algorithms. In this method, the weak learners are combined into a final sum representing the boosted output. We use the particular algorithm called AdaBoost-SAMME ~\cite{zhu2009multi} and use the decision trees as the base estimator. \\

\noindent  \textbf{Logistic Regression (LR)}. Logistic regression is a generalized linear model which uses a logistic function to infer the relationship between a dependent variable and one or more independent variables. We utilizes the binomial logistic regression which predicts the probability that an observation falls into one of the two categories. Logistic regression has low varience and is less prone to overfitting. \\

\noindent  \textbf{Naive Bayes Classifier (NB)}. Naive Bayes is a probabilistic classifier which is based on applying Bayes' theorem  with independence assumption between every feature pairs. Naive Bayes classifiers are highly scalable and less prone to the curse of dimensionality, making it one of the top machine learning algorithms. We implement the Gaussian Naive Bayes algorithm for classification where the likelihood of the features is assumed to be Gaussian.

\subsection{Measurement Group Comparison}
\begin{figure}[!h] 
	\centering
	\subfigure[]{%
		\includegraphics[width=1\columnwidth]{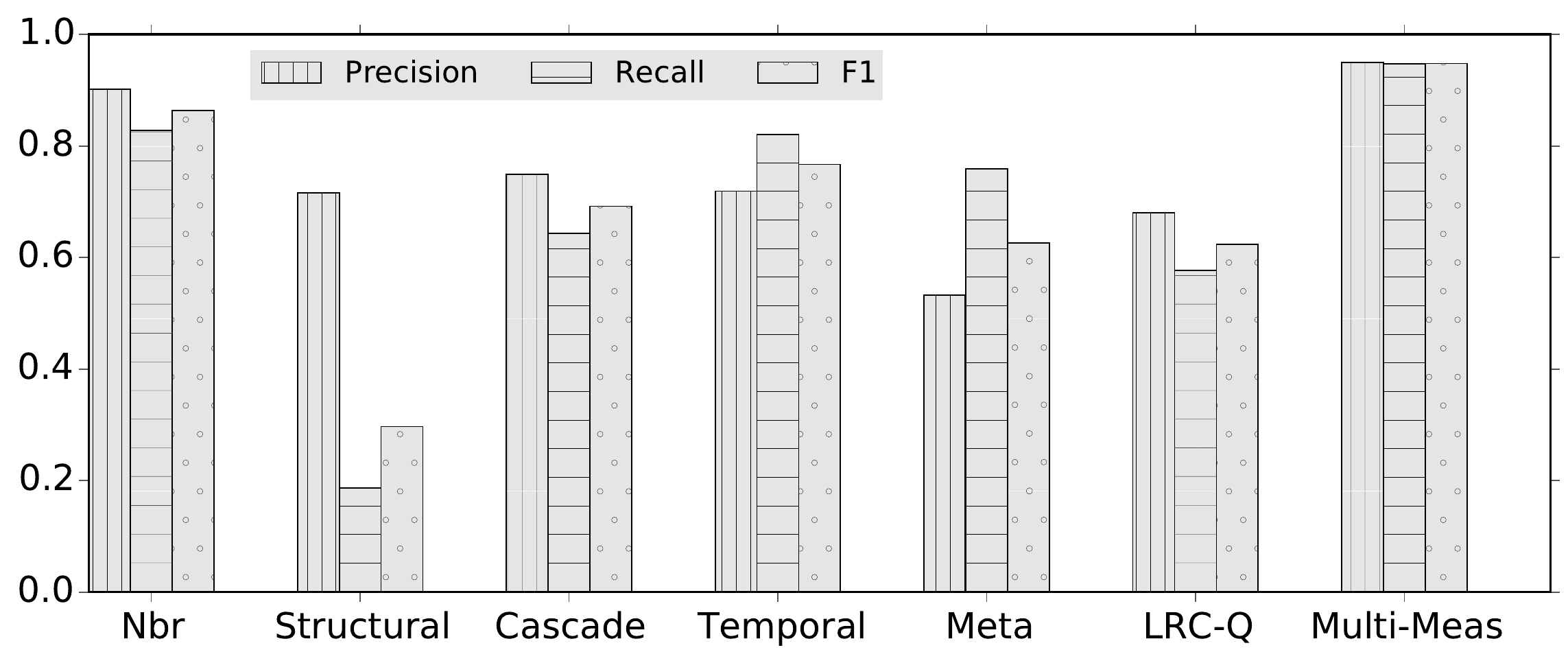}
		\label{fig:rf}
	}  
	\subfigure[]{%
		\includegraphics[width=1\columnwidth]{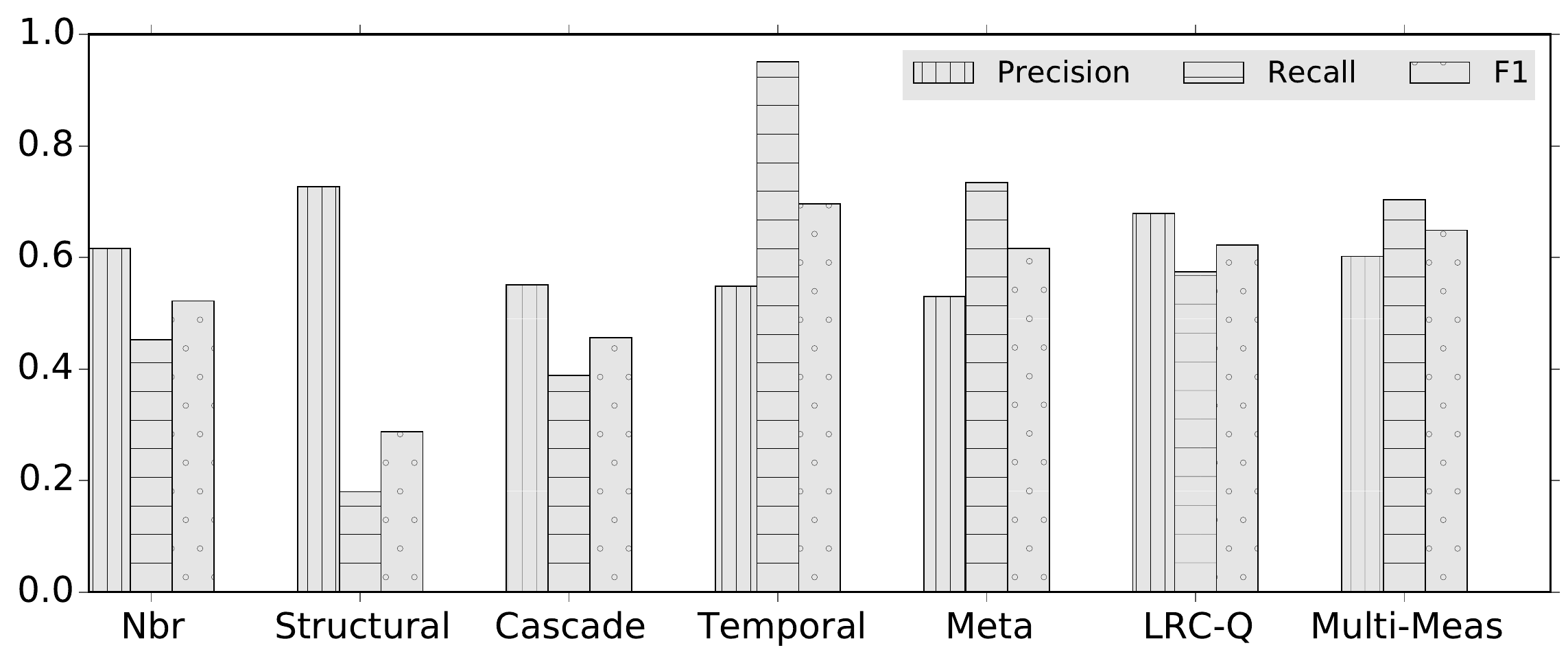}
		\label{fig:lr}
	} 
	\subfigure[]{%
		\includegraphics[width=1\columnwidth]{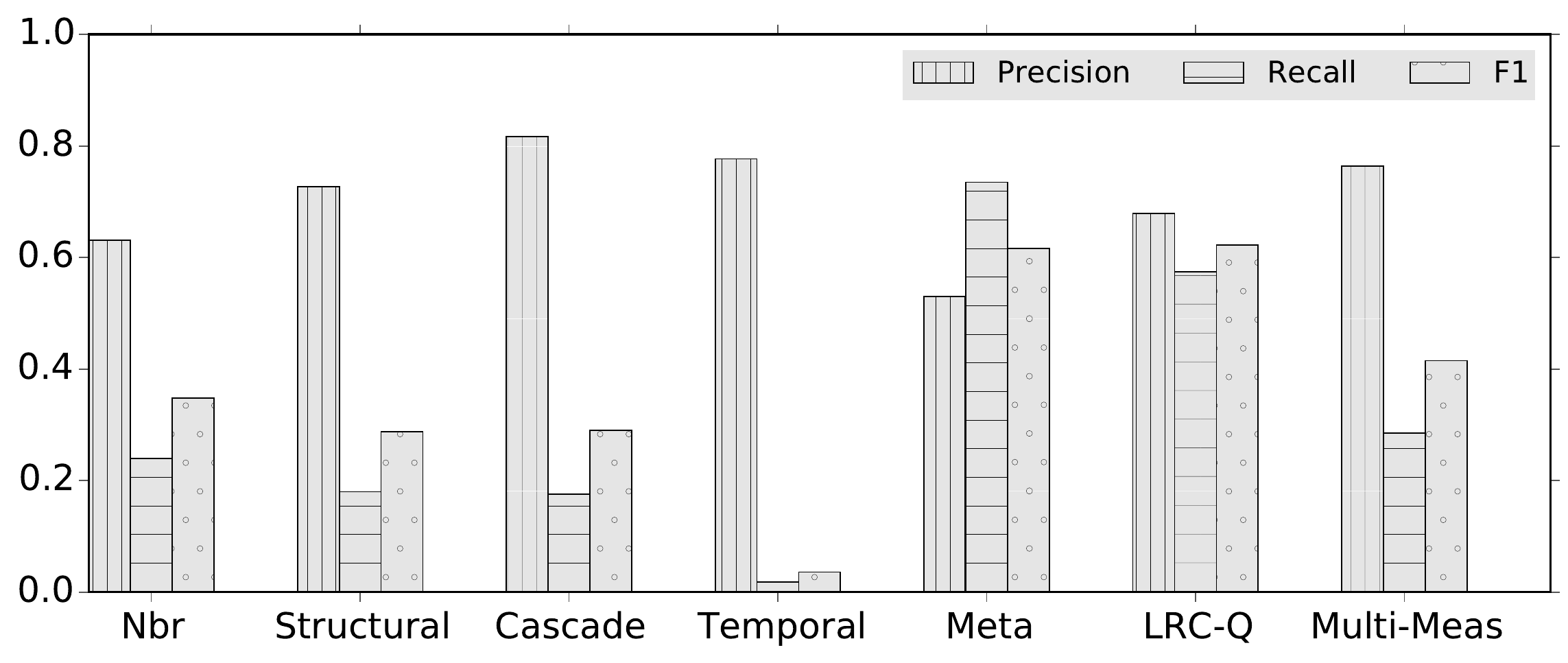}
		\label{fig:nb}
	}    
	\subfigure[]{%
		\includegraphics[width=1\columnwidth]{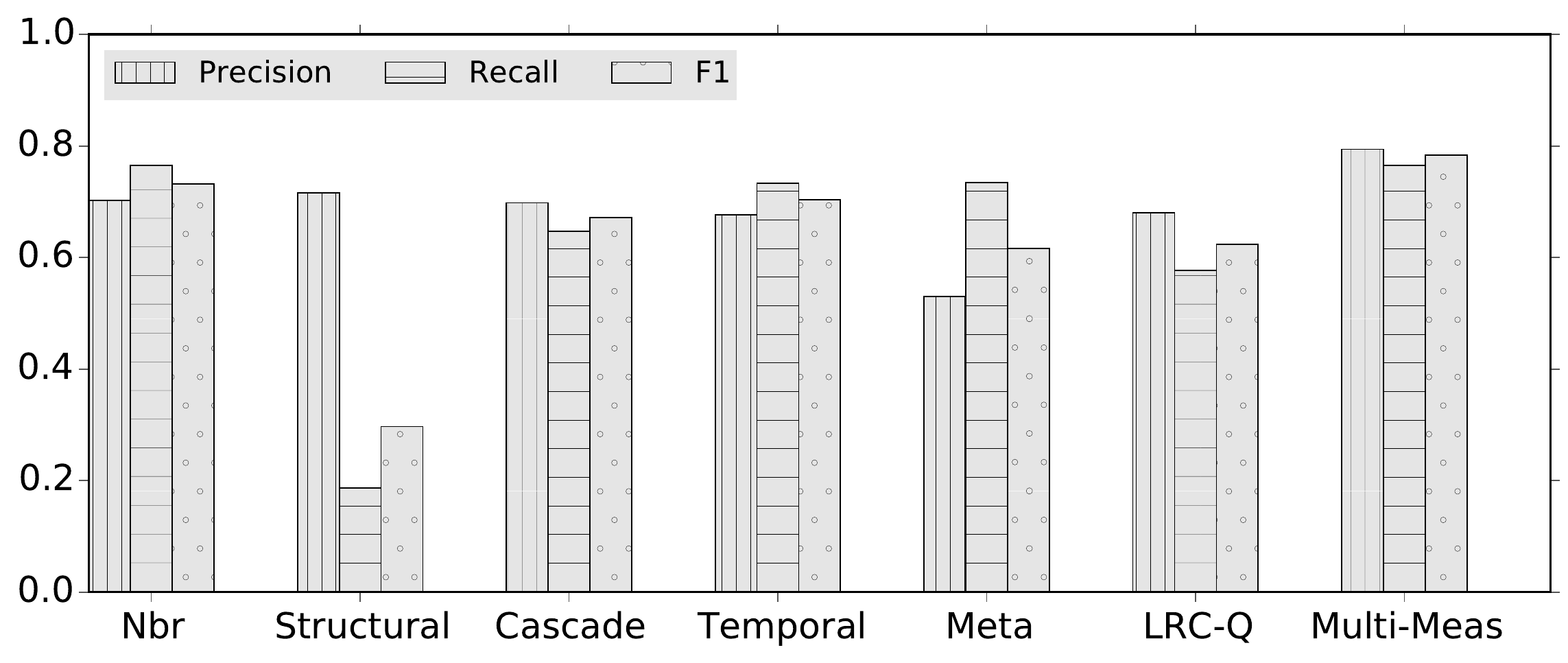}
		\label{fig:ab}
		
	}

	\caption{Performance with different classifier algorithms. a) Random Forest b) Logistic Regression c) Naive Bayes d) AdaBoost.}
	\label{fig:clf_grp}
\end{figure}

Here we compare the classification performance of the various measurement groups described in Section ~\ref{features}. Fig.~\ref{fig:clf_grp} shows the behavior of different feature groups using multiple classifier algorithms, which provides a better understanding of this all-important component in a deployed system. Generally Random Forest provides the best performance among all the classifier algorithms. Neighborhood-based (Nbr) measures perform quite well in Random Forest, AdaBoost and Logistic regression. This is consistent with what we discussed in Section~\ref{measStudy}. Structural diversity measures show less performance compared to other groups. This can be attributed to the fact that it is not often used independently in classification, and usually this group performs well in conjunction with other measures such as Neighborhood-based. LRC-Q gives performance measure comparable to the results in ~\cite{zhang2013social}. Cascade-based measures are observed to perform reasonably well in Random Forest, Logistic Regression and AdaBoost. This once again illustrates the significance of cascade size and brings into focus the path length measure. Temporal measure performs well in all classifiers except Naive Bayes. Although time based measures are frequently used as a decay factor in conjunction with other measures (\!\cite{goyal2010learning},~\cite{zhang2013social}), our results show that it could yield high predictive power by itself. Metadata measures show good and consistent performance across all classifiers. As research by ~\cite{jenders2013analyzing} shows, hashtag and mentions have high predictive power with respect to retweet behaviour and our results confirm the significance of this measure along with the hasLinks measure. 

With an eye toward a deployed system, we also examine a ``Multi-Measurement model'' which is a combination of Neighborhood, Structural, Cascade, Temporal and Metadata measures. The Multi-Measurement model shows better performance than individual groups generally among Random Forest, Logistic Regression and AdaBoost classifiers. The other measures such as neighborhood-based, temporal and LRC-Q perform reasonably well compared to rest of the individual future groups. The performance of Multi-Measurement model shows real value in combining the various features and individual feature groups to improve our ability to predict retweet behavior in real world datasets.

\subsection{Multi-Measurement Model Compared to Influence Locality}

We compare our results with the LRC-Q model described in \cite{zhang2013social}.  We experimented with multiple classification algorithms for this task and the best results were obtained using Random Forest classifier. The results obtained using Random Forest (RF), Logistic Regression (LR), Naive Bayes (NB) and AdaBoost (AB) are shown in the Table~\ref{tab:clf_res}. As LRC-Q uses only a single feature, we only use Logistic Regression for its evaluation. It can be observed that  Multi-Measurement model outperforms the LRC-Q model in all classifiers except for Naive Bayes. This can be attributed to the fact that while LRC-Q takes into account pairwise and structural influence along with time decay, Multi-Measurement model incorporates more parameters in addition to the above. LRC-Q has combined the pairwise and structural factor into a single feature and uses time measure as a decay factor. The Multi-Measurement model on the other hand treats them individually, along with including different kinds of pairwise influence (such as active neighbor count, personal network exposure and average in-neighbors of active neighbors), considering both direct as well as ratio based measures for structural diversity, and using temporal measure as an independent feature. In addition to that, this model also includes cascade and metadata based features giving it a broader view of the parameters that can influence an individual's retweeting behavior. This demonstrates that in any attempt of retweet prediction, a broader approach is required, which incorporates multiple measures that are are closely related (within the measurement groups) and those that are mutually exclusive (across groups) to obtain the best prediction in classification. \\

\begin{table}[!h]
	\centering
	\small
	\begin{tabular}{|c|c|c|c|} \hline
		Model & Precision & Recall & F1 \\ \hline
		LRC-Q (LR) & 0.679 & 0.573 & 0.622 \\ \hline
		Multi-Meas (RF) & \textbf{0.95} & \textbf{0.947}  & \textbf{0.948} \\ \hline
		Multi-Meas (AB) & 0.794 & 0.765  & 0.784 \\ \hline
		Multi-Meas (LR) & 0.602 & 0.704  & 0.649 \\\hline
		Multi-Meas (NB) & 0.764 & 0.285  & 0.415 \\ \hline
		
	\end{tabular}
	\caption{Performance of retweet behavior prediction}
	\label{tab:clf_res}
\end{table}

\section{Varying Negative to Positive ratio}
\label{imbalSec}

\vspace{5pt}

\begin{figure}[!h]  
	\centering
	\subfigure[]{%
		\includegraphics[width=.52\columnwidth]{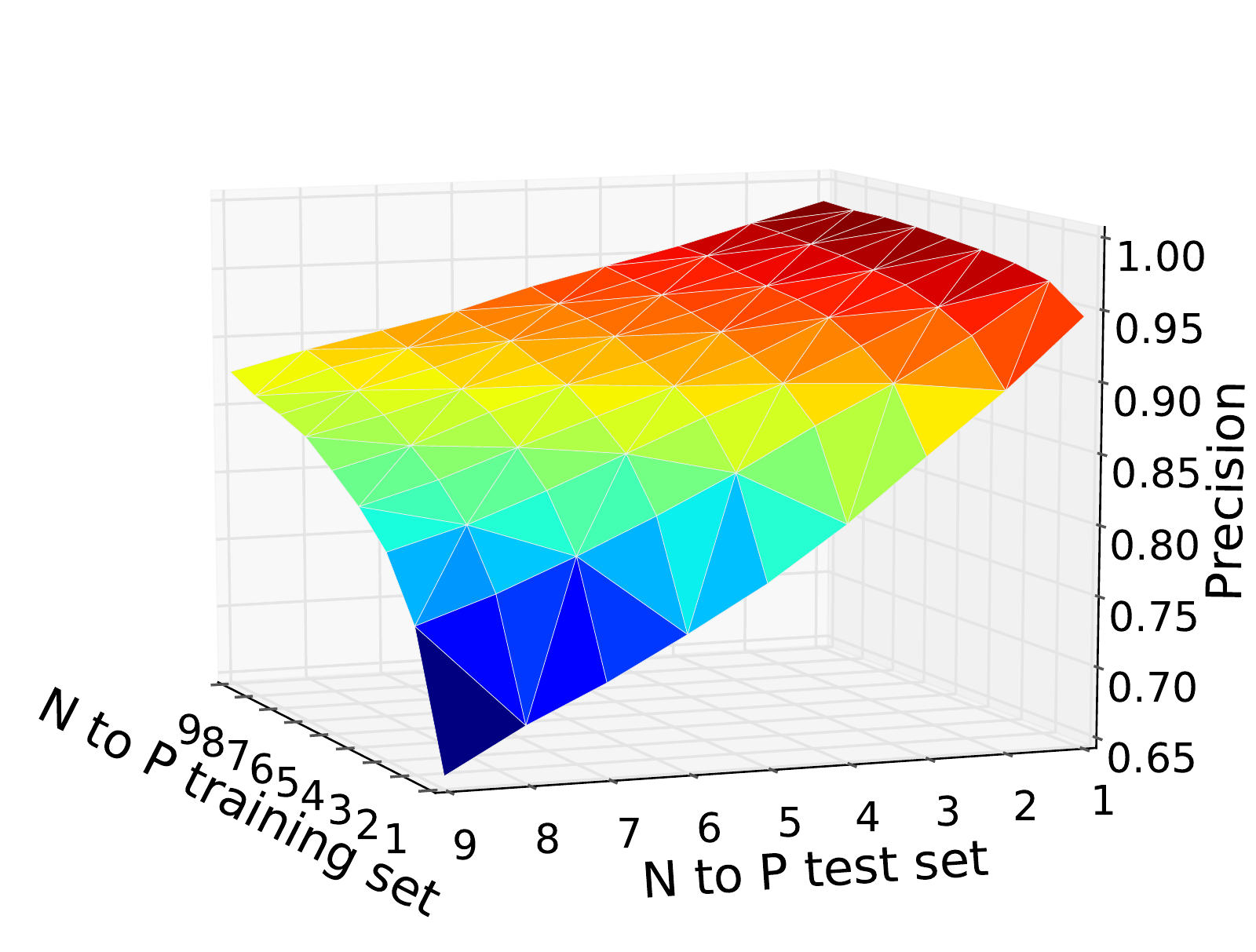}
		\label{fig:prec3d}
	} 
	\subfigure[]{%
		\raisebox{.08\height}{ 
			\includegraphics[width=.425\columnwidth]{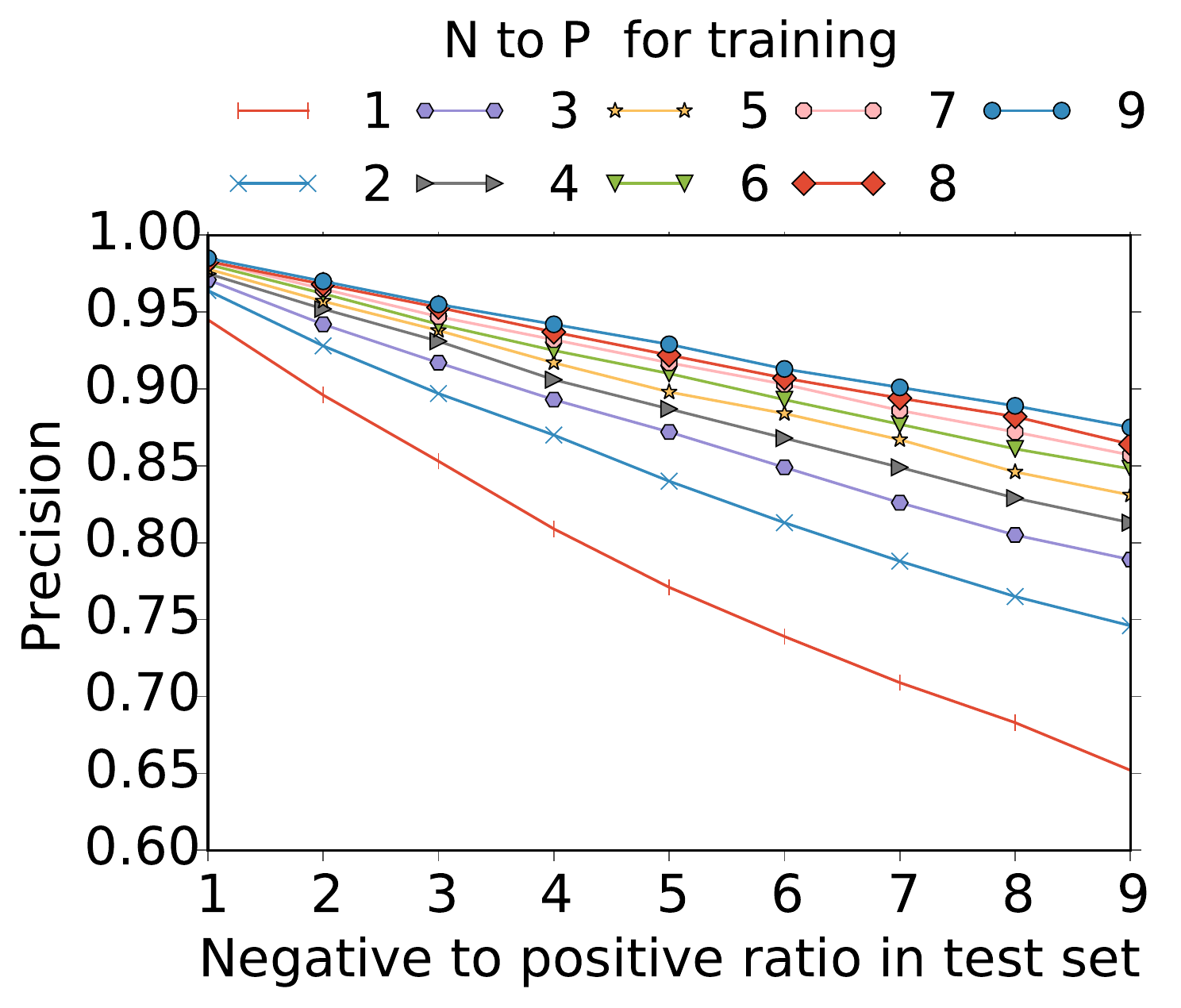}}
		\label{fig:prec2d}
	} 
	\subfigure[]{%
		\includegraphics[width=.52\columnwidth]{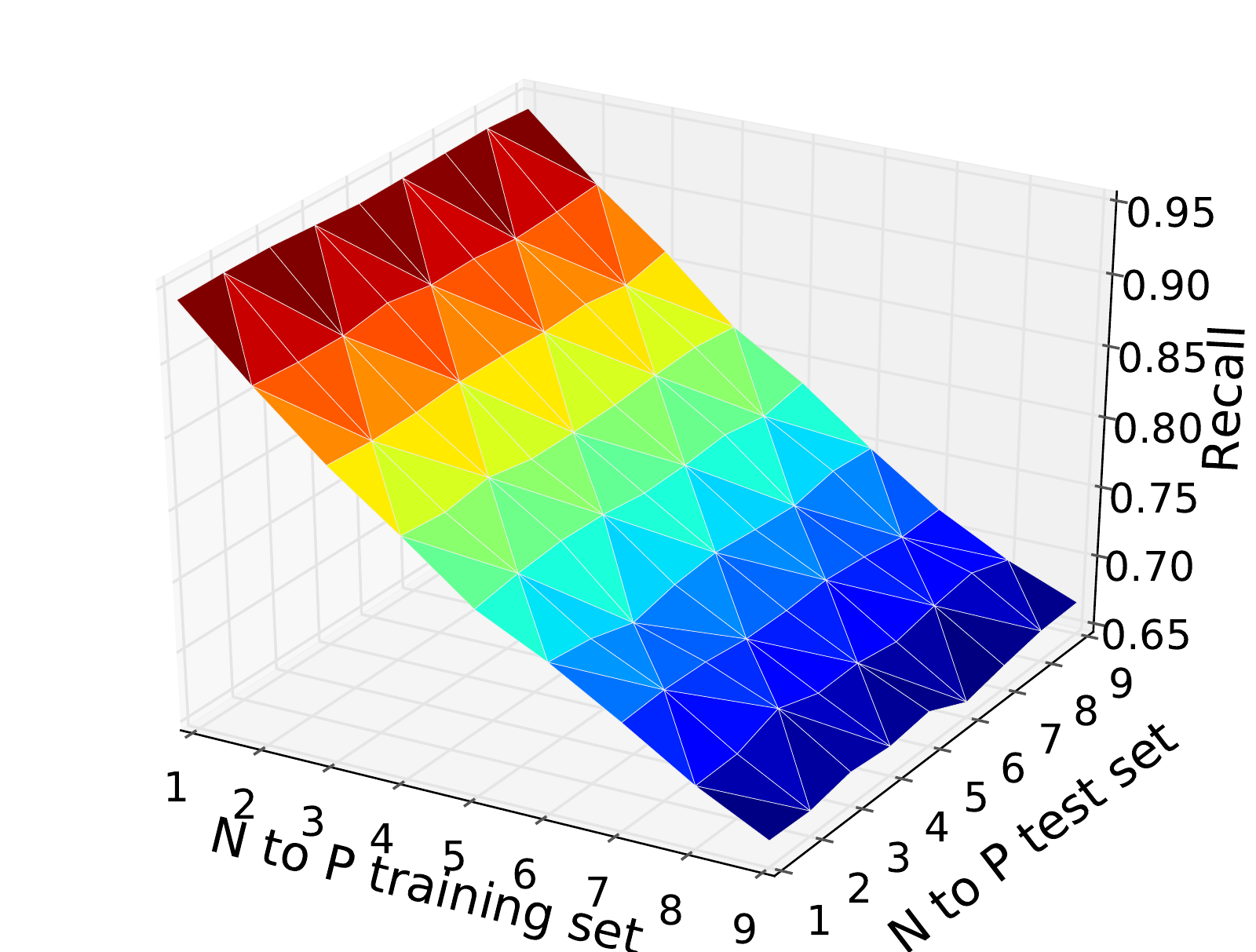}
		\label{fig:rec3d}
	} 
	\subfigure[]{%
		\raisebox{.07\height}{ 
			\includegraphics[width=.425\columnwidth]{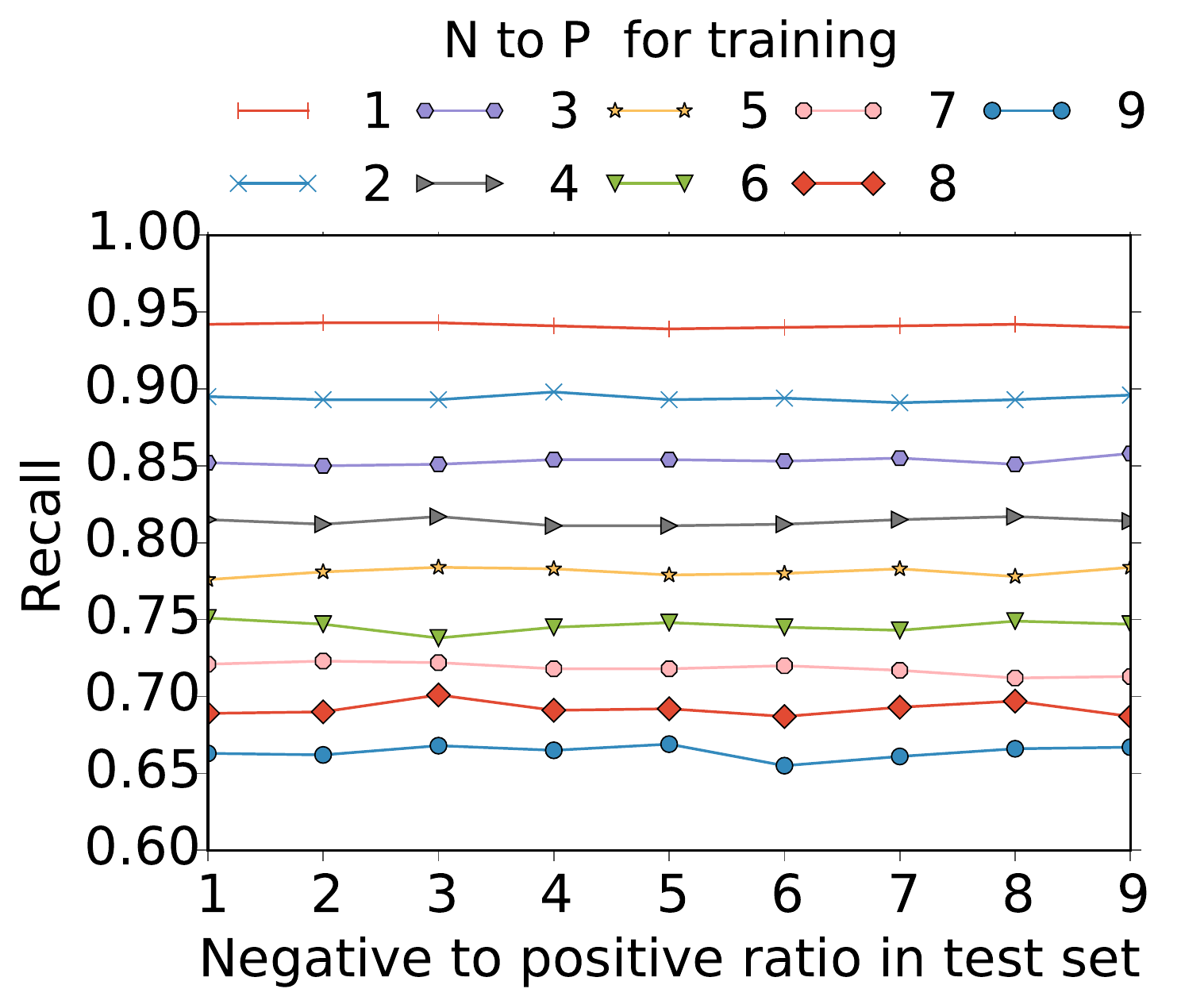}}
		\label{fig:rec2d}
	} 
	\subfigure[]{%
		\includegraphics[width=.52\columnwidth]{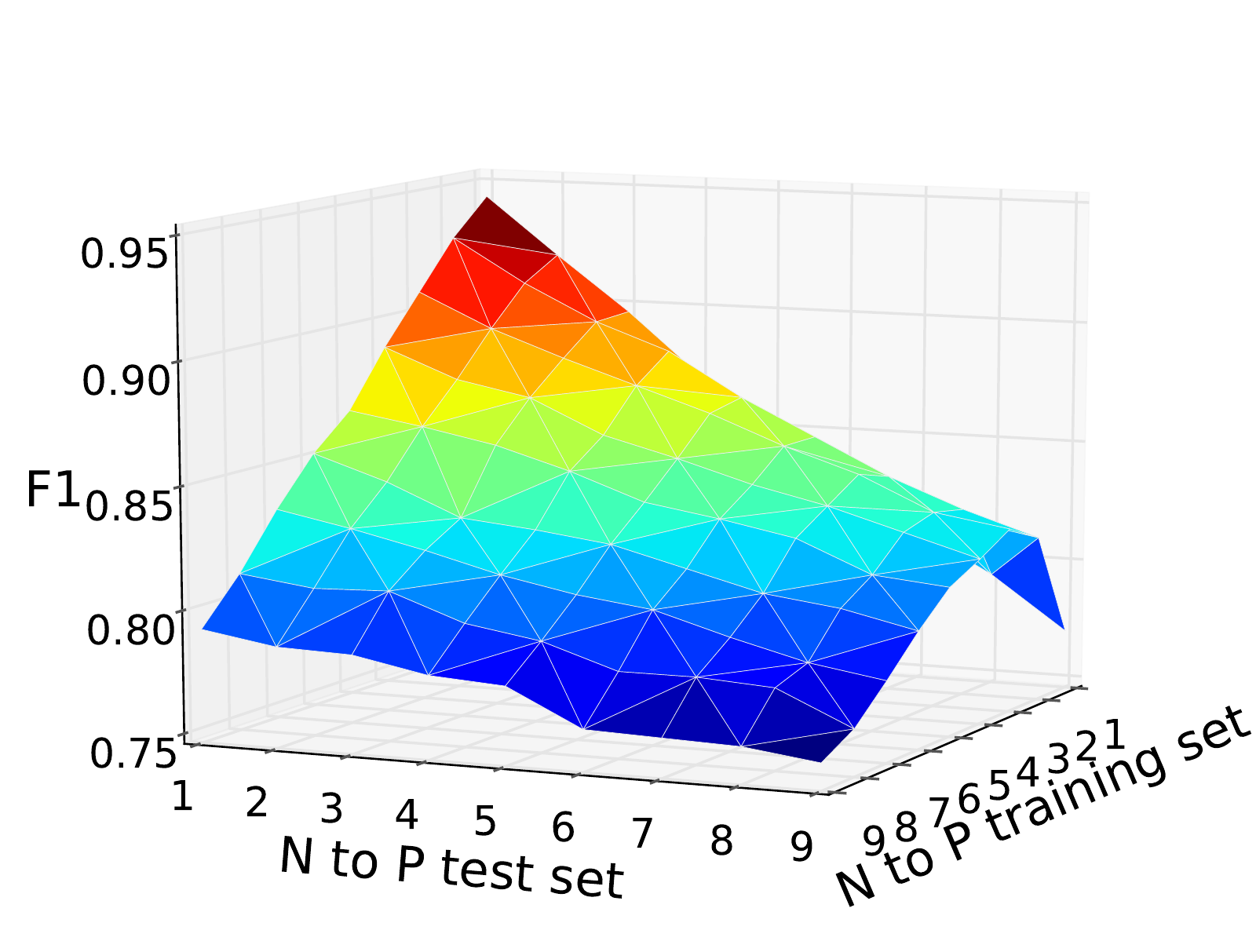}
		\label{fig:f13d}
		
	}   
	\subfigure[]{%
		\raisebox{.08\height}{ 
			\includegraphics[width=.425\columnwidth]{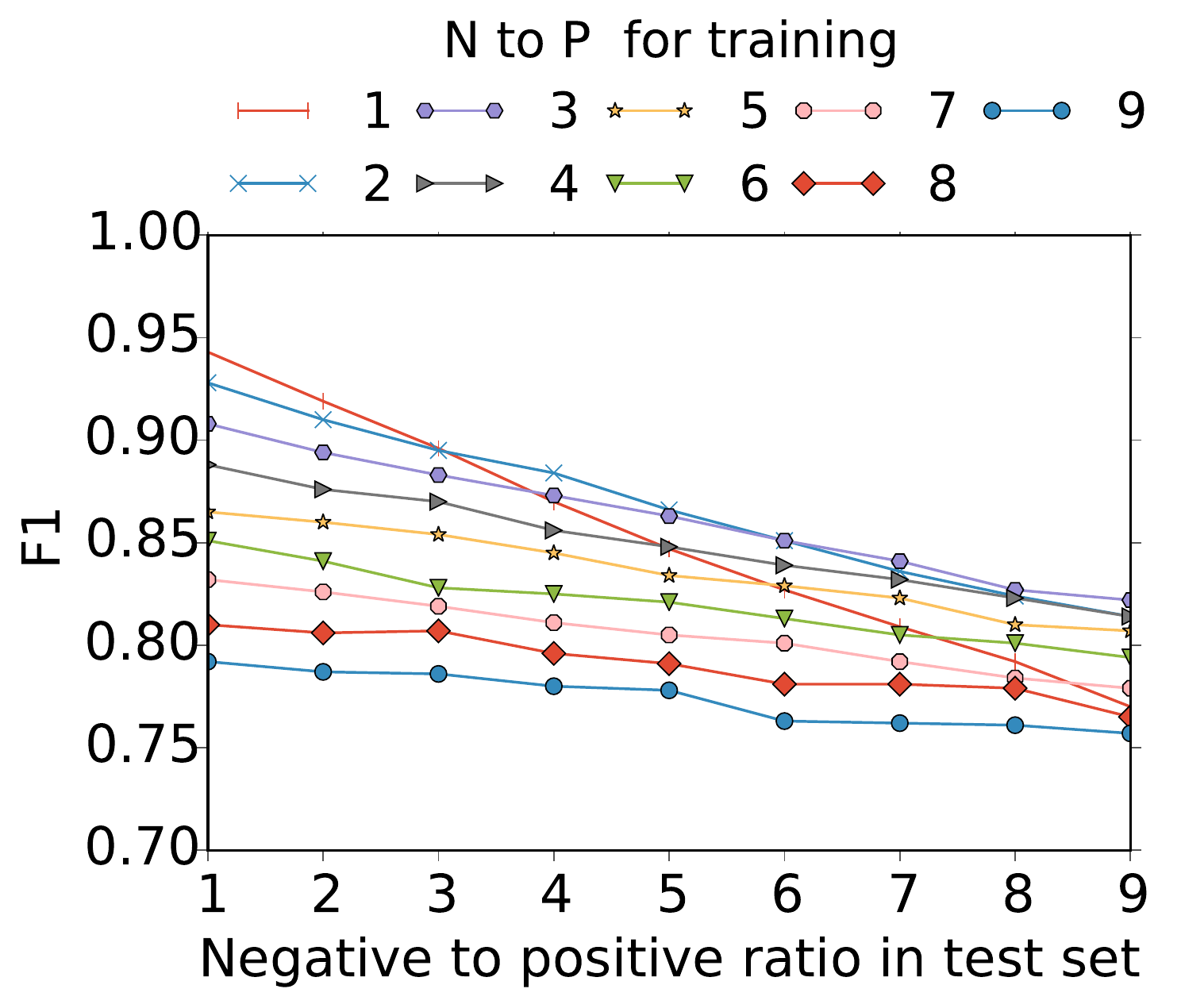}}
		\label{fig:f12d}
	} 
	\caption{Plots for classification on imbalanced data for Multi-Measurement model using Random Forest. a) Precision surface plot b) Precision line plot c) Recall surface plot d) Recall line plot e) F1 surface plot f) F1 line plot.}
	\label{fig:3dplot}
\end{figure}

An important question when deploying the aforementioned methods in a real-world application is how to best train the model to cope with data imbalance observed in-practice.  As individuals are exposed to an arbitrarily large number of microblogs that they do not rebroadcast, this is a difficult - and unfortunately relatively unstudied problem. Here, we conducted experiments to analyse how classification performance varies with different negative to positive ratio in both training and test set. The surface and linear plots in Fig.~\ref{fig:3dplot} show the precision, recall and F1 values obtained using Random Forest classifier, when negative to positive ratio is varied from 1:1 to 9:1. The ratio was varied in both training set and test set to observe the effects on overall performance. Precision is observed to decrease as we increase the size of negative samples in test set while keeping the ratio in training set constant. Recall is observed to remain the same with changing ratio in test set. Change in negative to positive ratio in training set on the hand, shows slight increase in precision where as recall decreases. Results for LRC-Q follows a similar pattern except for the convergence of recall for increased imbalance in training set. From these results, it can be generally observed that 1:1 is the ideal ratio of negative to positive samples in training set for an unknown imbalance in test data.

\begin{figure}[!h]  
	\centering
	\subfigure[]{
		\includegraphics[width=.50\columnwidth]{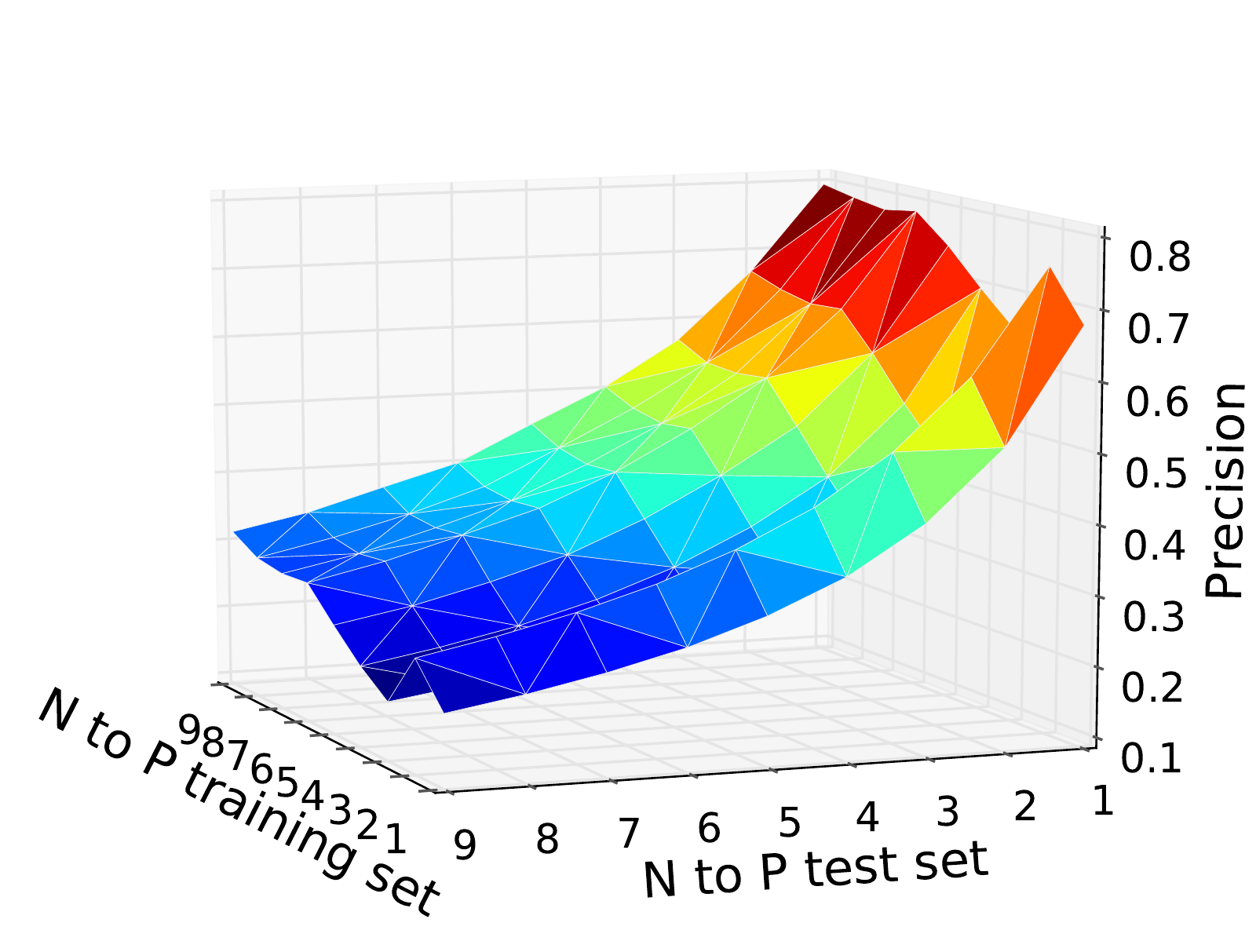}
		\label{fig:prec3dLr}
	} 
	\subfigure[]{%
		\includegraphics[width=.425\columnwidth]{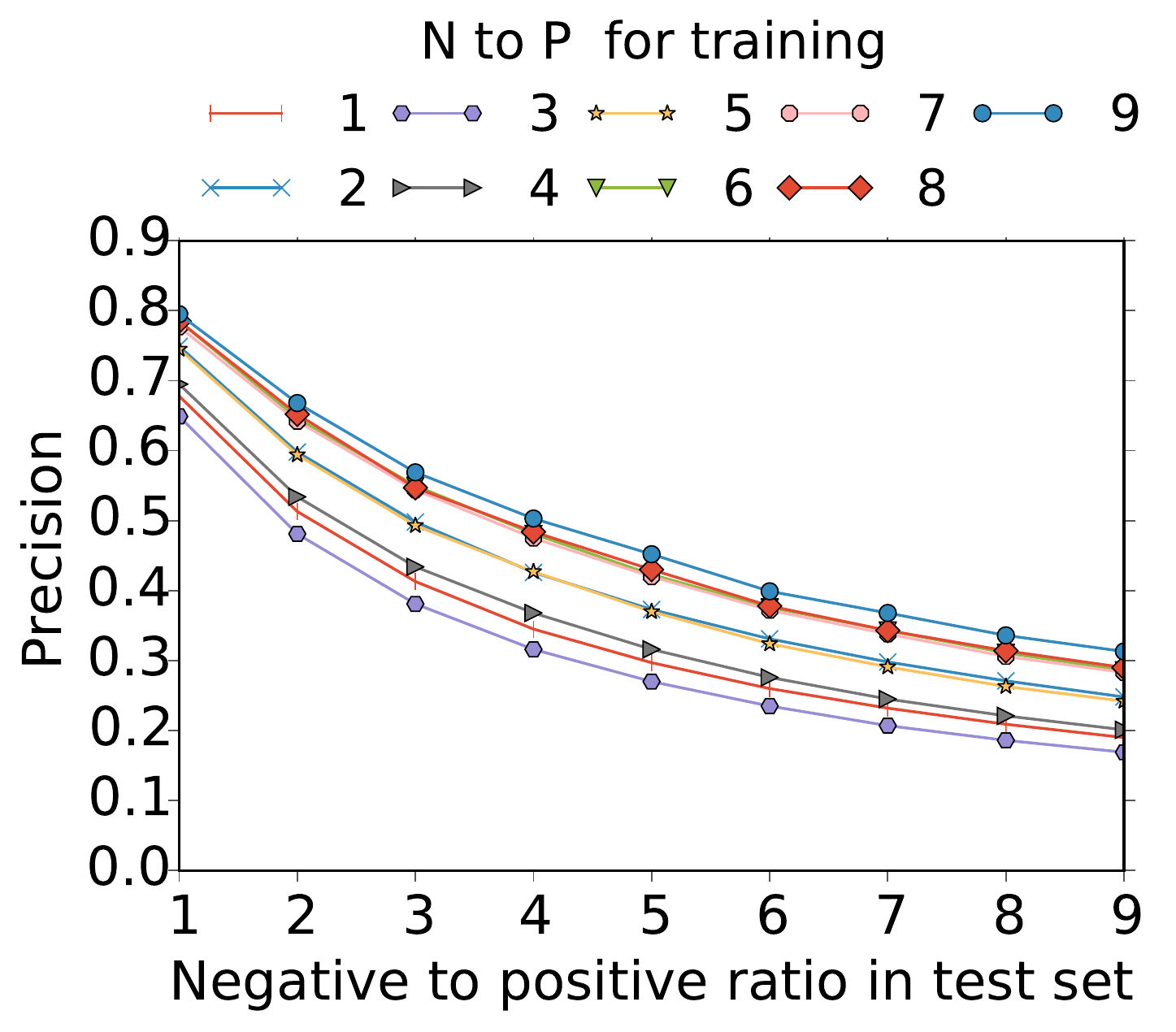}}
	\label{fig:prec2d}
	\subfigure[]{
		\includegraphics[width=.50\columnwidth]{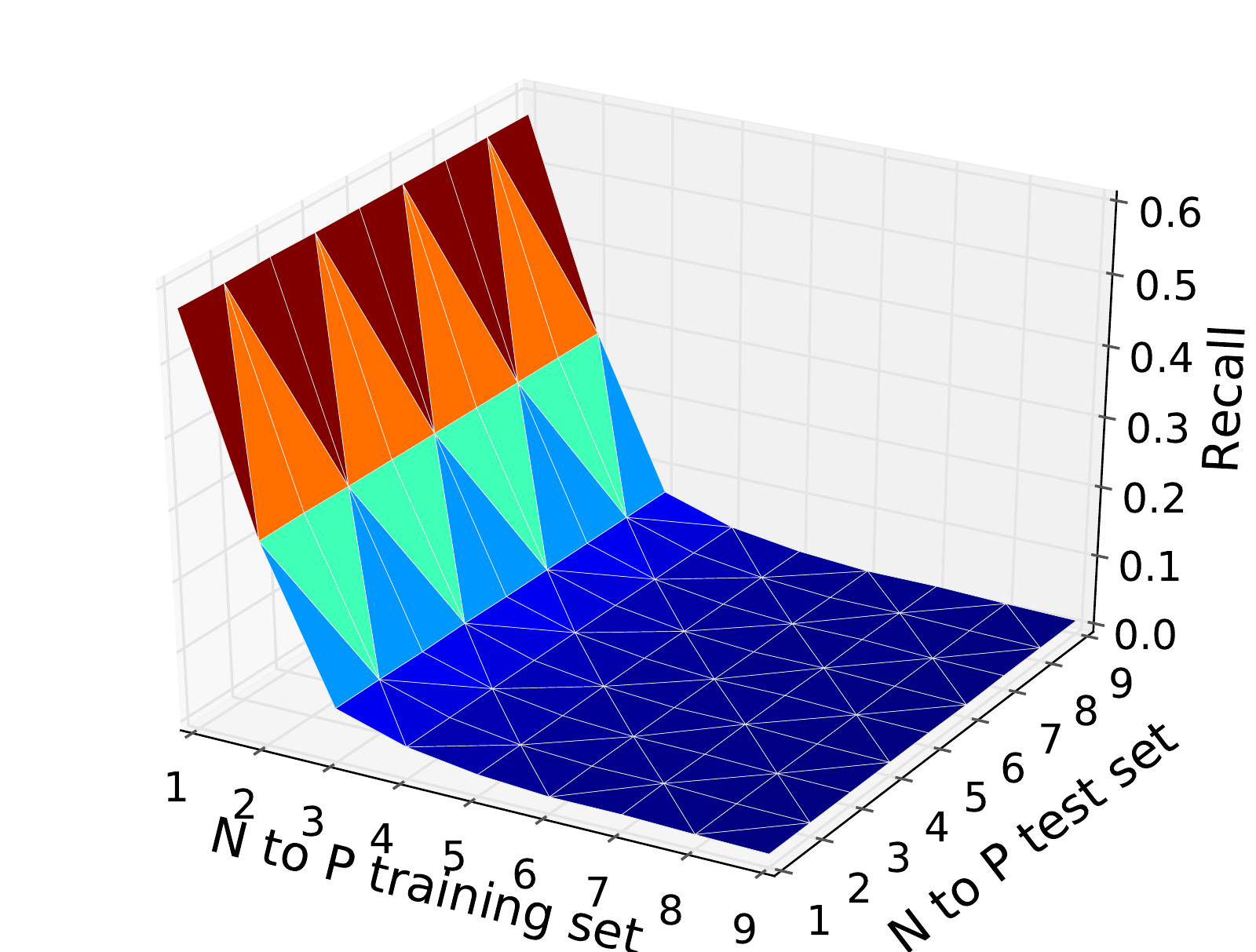}
		\label{fig:rec3dLr}
	} 
	\subfigure[]{%
		\raisebox{.1\height}{ 
			\includegraphics[width=.425\columnwidth]{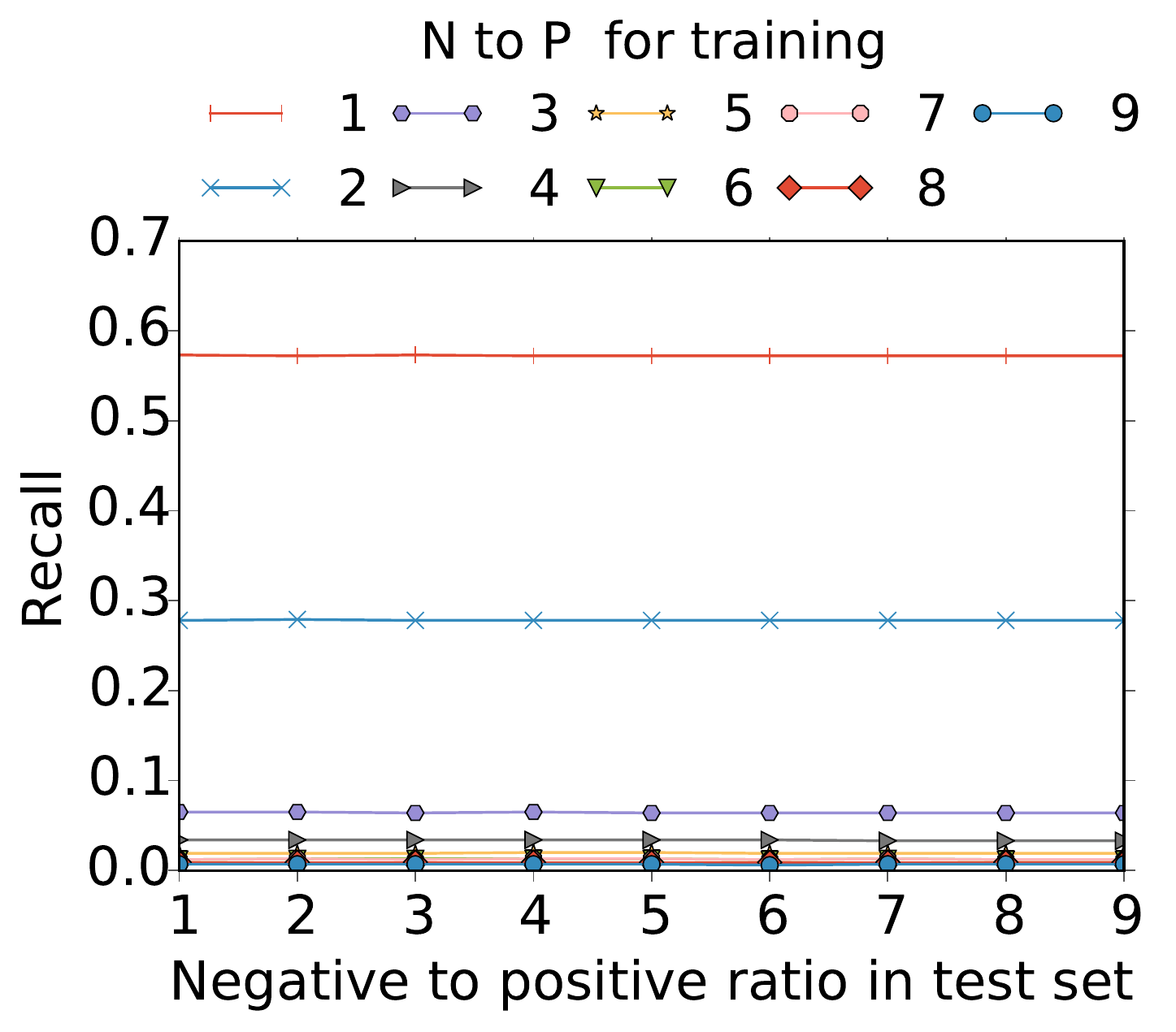}}
		\label{fig:rec2dLr}
	} 
	\subfigure[]{
		\includegraphics[width=.50\columnwidth]{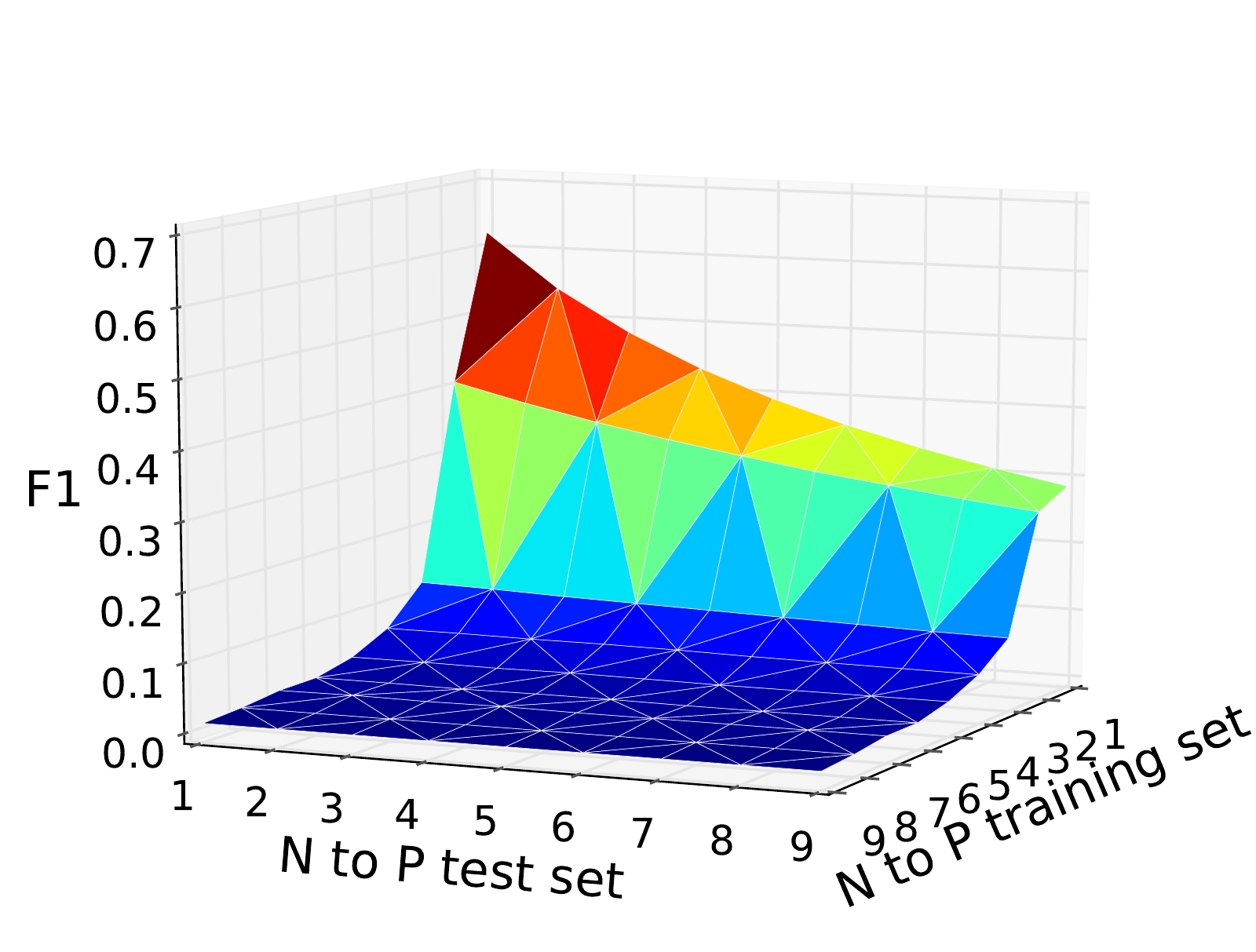}
		\label{fig:f13dLr}
	}    
	\subfigure[]{%
		\raisebox{.1\height}{ 
			\includegraphics[width=.425\columnwidth]{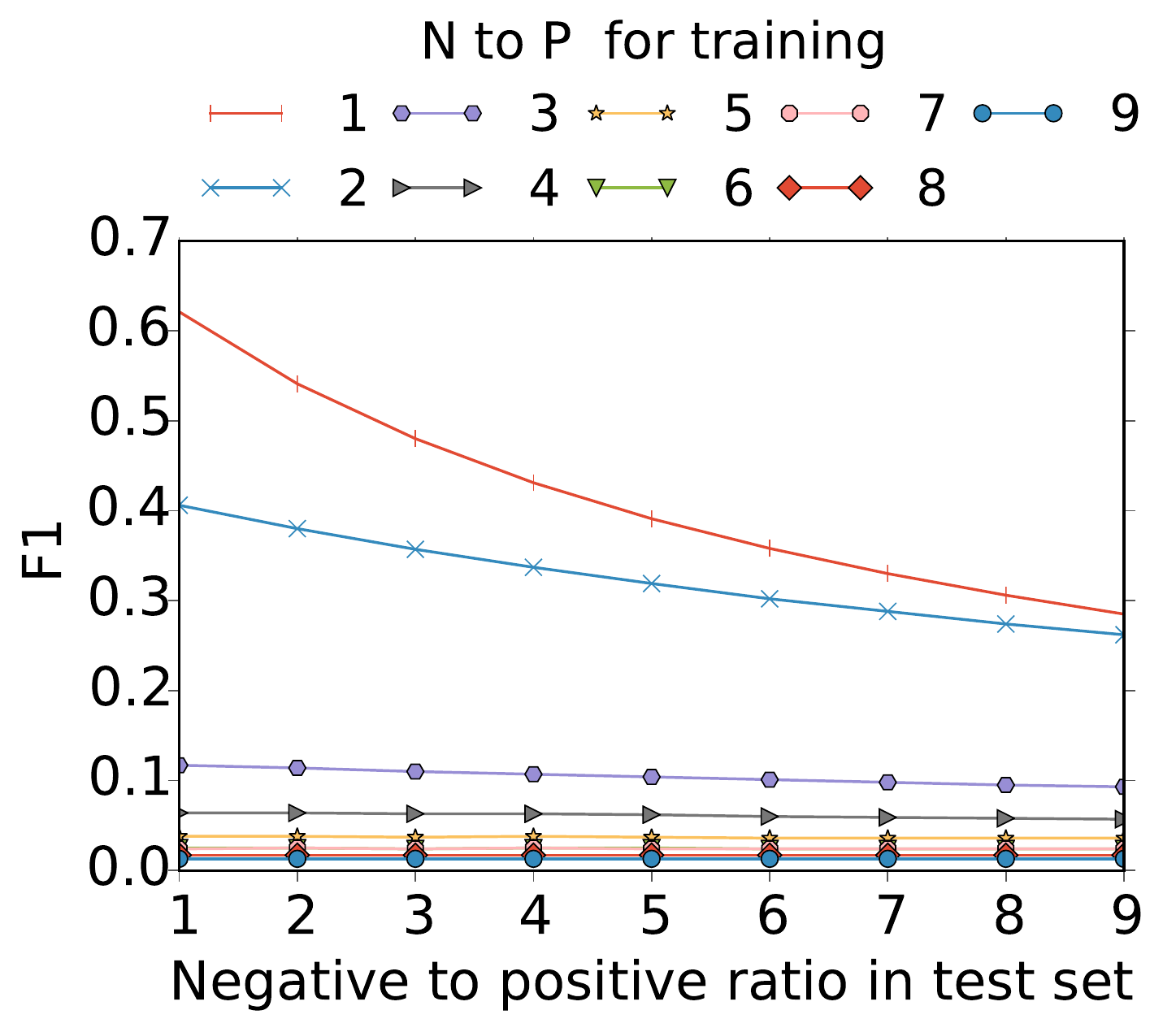}}
		\label{fig:f12dLr}
		
	} 
	\caption{Plots for classification on imbalanced data for LRC-Q using Logistic Regression. a) Precision surface plot b) Precision line plot c) Recall surface plot d) Recall line plot e) F1 surface plot f) F1 line plot.}
	\label{fig:3dplotLr}
\end{figure}

\section{Conclusion}

In this paper, we examines the performance of a wide variety of social network based measurements and study the probability of an individual becoming influenced based on them. In this study, we grouped those measures under various measurement groups to understand their group wise predictive power. We designed these experiments so that they would move beyond standard research-based experiments used to evaluate an idea - we designed these experiments to understand how well these ideas can be used in a deployed system. We look to use these results in a system that we intend to deploy or license for real-world influence operations such as counter-extremism.

\section*{Acknowledgments}

Some of the authors are supported through the AFOSR Young Investigator Program (YIP) grant FA9550-15-1-0159, ARO grant W911NF-15-1-0282, the DoD Minerva program grant N00014-16-1-2015 and the EU RISE program.

\bibliographystyle{IEEEtran}
\bibliography{ref}
  
\end{document}